\documentclass[compsoc,conference,a4paper,10pt,times]{IEEEtran}
\IEEEoverridecommandlockouts

\usepackage[english]{babel}
\usepackage[letterpaper,top=2cm,bottom=2cm,left=3cm,right=3cm,marginparwidth=1.75cm]{geometry}

\usepackage{amsmath}
\usepackage{graphicx}
\usepackage[colorlinks=true, allcolors=blue]{hyperref}
\usepackage{soul}
\usepackage{xurl}
\usepackage{longtable}
\usepackage{booktabs, tabularx, multicol, multirow}
\usepackage{array} 
\usepackage{geometry}
\usepackage{lipsum}
\usepackage{color, colortbl}
\usepackage[numbers, sort]{natbib}
\usepackage{mwe}
\usepackage{tikz}
\usetikzlibrary{shapes.geometric, arrows.meta, positioning, backgrounds, fit, calc}
\definecolor{lightgray}{rgb}{0.92, 0.92, 0.92}
\usepackage{enumitem}
\setlist{leftmargin=1em}

\title{SoK: Honeypots \& LLMs, More Than the Sum of Their Parts?
}

\author{
\IEEEauthorblockN{1\textsuperscript{st} Robert A. Bridges}
    \IEEEauthorblockA{
        \textit{AI Labs} \\
        \textit{AI Sweden}\\
        Gothenburg, Sweden \\
        robert.bridges@ai.se
        }

\and
\IEEEauthorblockN{2\textsuperscript{nd} Thomas R. Mitchell}
    \IEEEauthorblockA{\textit{Security R\&D} \\
        \textit{Volvo Group}\\
        Gothenburg, Sweden \\
        thomas.mitchell@volvo.com
        }
\and
\IEEEauthorblockN{3\textsuperscript{st} Mauricio Muñoz}
    \IEEEauthorblockA{
        \textit{AI Labs} \\
        \textit{AI Sweden}\\
        Gothenburg, Sweden \\
        mauricio.munoz@ai.se
        }
\and
\IEEEauthorblockN{4\textsuperscript{th} Ted Henriksson}
    \IEEEauthorblockA{
        \textit{AI Labs} \\
        \textit{AI Sweden}\\
        Gothenburg, Sweden \\
        ted.henriksson@ai.se
        }

}
\begin{document}
\maketitle

\thispagestyle{plain}\pagestyle{plain}

\begin{abstract}
The advent of Large Language Models (LLMs) promised to resolve the long-standing paradox in honeypot design: achieving high-fidelity deception with low operational risk.
Since late 2022, a flurry of research has demonstrated steady progress from ideation to prototype implementation.
While promising, evaluations show only incremental progress in real-world deployments, and the field still lacks a cohesive understanding of emerging architectural patterns, core challenges, and evaluation paradigms.
To fill this gap, we provide the first comprehensive overview and analysis of this new domain, focusing on three critical, intersecting research areas:
we provide a taxonomy of honeypot detection vectors, mapped to how LLM-based simulation can or cannot aid deception;
we synthesize the emerging literature on LLM-powered honeypots, identifying a canonical architecture, an evaluation tetrad, and an attacker trichotomy mapped to honeypot requirements; 
and we chart the evolution of honeypot log analysis into automated intelligence generation.
Finally, we synthesize these findings into a forward-looking research roadmap, 
arguing that the true potential of this technology lies in creating autonomous, self-improving deception systems to counter the emerging threat of intelligent, automated attackers.
\end{abstract}



\section{Introduction}\label{sec:intro}
Cybersecurity honeypots---decoy systems designed to lure and analyze attackers---have long grappled with a fundamental design paradox.
The spectrum of honeypots ranges from ``low-interaction'' systems, which are safe and easy to deploy but simplistic and easily detected, to ``high-interaction'' systems, which offer high-fidelity engagement at the cost of significant risk and management overhead.
For decades, the central goal of honeypot research has been to resolve this tension: to achieve the high fidelity of a real system with the low risk and minimal effort of a simulated one.

The recent advent of Large Language Models (LLMs) appeared to offer a silver bullet.
With their uncanny ability to generate convincing text, code, and shell interactions, LLMs promised a paradigm shift for honeypots.
The vision was clear: low-interaction decoys could finally be imbued with dynamic, adaptive, and believable interaction profiles, luring attackers into revealing their techniques.\footnote{While our focus is on enterprise network server emulation, researchers have explored other creative applications, including honeypots for IoT devices~\cite{vasilatos2024llmpot}, systems to misdirect web scrapers~\cite{tatoris2025}, and chatbots designed to waste the resources of email and phone scammers~\cite{cambiaso2023scamming, AIDaisyGranny}.}

This vision spurred a flurry of research seeking to integrate LLMs into existing honeypot frameworks.
As this work was often performed concurrently, many of these formative papers did not cite or build directly upon each other.
Standardized metrics and benchmarks are emerging but have yet to achieve widespread adoption; consequently, rigorous comparative analysis remains limited.
A comprehensive understanding of how this new field is developing is needed to identify its fundamental limitations, direct research to the most promising or most needed directions, and mitigate the fragmentation inherent in rapidly evolving research areas.

We provide the first systematic review of LLM-powered honeypots, uniquely situating it between surveys of two critical, flanking topics: honeypot fingerprinting, to understand the core problems of realism, and honeypot data analysis, to understand the ultimate goal of intelligence generation.
The goal of this Systematization of Knowledge (SoK) paper is to provide the necessary foundational understanding of this burgeoning research area and to accelerate the development of effective LLM-powered defenses.

This paper is organized as follows.
Section \ref{sec:background} provides background on cybersecurity and honeypots. 
Section \ref{sec:surveys} contains our three core surveys.
Section \ref{sec:discussion} presents our synthesized takeaways and \ref{sec:future-research} is a forward-looking research roadmap.

\subsection{Related Works \& Our Contributions}
\label{sec:related-works}
To situate our work, we conducted a review of recent surveys and systematization of knowledge papers.
Our search on Google Scholar for the terms \texttt{("survey" OR "systematization of knowledge" OR "SoK") AND "honeypot" AND ("Large Language Model" OR "LLM")} from 2024 to the present yielded no prior SoKs on this specific topic.
The closest works are broader surveys on the impact of LLMs on cybersecurity~\cite{hasanov2024application, hassanin2024comprehensive, guo2025frontier}.
While these papers acknowledge honeypots as a potential application, they are discussed only briefly within a much larger context.
Our work is therefore complementary, providing a deep and focused systematization solely on this intersection.
The most relevant prior survey on modern deception technology is by Javadpour et al.~\cite{javadpour2024comprehensive}.
This comprehensive work offers a  review of cyber deception literature from 2008-2023, establishing a clear baseline for the pre-LLM era.

\subsubsection*{Contributions}
Our work distinguishes itself from these prior surveys not only through its focused scope but by providing several novel systematizations that structure this emerging field.
While Javadpour et al. provide a definitive look at the classic era of deception technology, our work provides the first systematic treatment of the LLM-driven era. 
\begin{itemize}
    \item We synthesize honeypot detection research into {four persistent fingerprinting vectors}. This framework delineates the capabilities LLM simulation can realistically advance versus those it fundamentally cannot address.
    \item We systematize the literature on LLM-powered honeypots by: defining an emerging {canonical honeypot architecture}; 
    defining an {attack model trichotomy} (Tier 1: Scripted Bots, Tier 2: LLM-Agents, Tier 3: Humans) to understand honeypot requirements; 
    introduce an {evaluation tetrad}---believability, fidelity, attacker cost, and defender cost---to enable rigorous comparative analysis and map it to the fingerprinting vectors;
    and characterizing the {``data desert''}, only having low-sophistication traffic and user tests hinders iterative development and obscures LLM efficacy.
    \item We {systematize honeypot log analysis}, charting the ``semantic leap'' from simple data reduction to the automated, agent-driven operationalization.
    \item Finally, we synthesize these findings into a forward-looking {research roadmap}, identifying high-impact directions such as the need for dynamic evaluation ecosystems and autonomous, self-improving feedback loops.
\end{itemize}
By providing these frameworks, our SoK offers a structured, multi-faceted understanding of the challenges, architectures, and future trajectory of LLM-powered honeypots, a contribution distinct from both the broad cybersecurity surveys and the pre-LLM deception literature.
We acknowledge the relative immaturity of the LLM-honeypot field; however, we maintain that early systematization is an overall boon, even if a later SoK is warranted, as it provides self-reflection and a guiding light for research community organization and scientific progress. 
    \subsection{Summarized Findings}
Historical honeypot fingerprinting principles often still apply, and we itemize these in Table \ref{tab:honeypot-vectors-llm}. 
Even with LLM simulation, fundamental limitations will give these systems away to an intelligent adversary. 
How to build robust deterministic responses and logic for when to use LLM simulation can assist in believability, cost, and security, but brings us to budding architectural needs for LLM honeypots

Architecturally, we identify an emerging canonical model for LLM-honeypots (Figure \ref{fig:architecture}) that incorporates components for filtering, state management, and generation, but we note a significant gap in open-source tooling.
We systematize evaluations of LLM honeypots in the research into a four-fold set of  criteria---believability, fidelity, attacker costs/intelligence, defender cost---each with different methodologies. 
This provides a framework for the ``what'' and ``how'' to measure honeypot performance, and lays the groundwork for comparison and optimization. 

Importantly, there persists an evaluation paradox: while  methodologies exist, their utility is hampered by a real-world ``data desert". 
Most studies use real-world deployments generating high volumes of low-sophistication attack traffic, for which LLM simulation is likely too costly and sophisticated or small human user studies, an attacker too smart for LLM simulation alone to be convincing. 
This leads to our finding that the appropriate target for these systems is neither pre-scripted bots nor skilled humans, but the emerging threat of sophisticated automated agents. 
Against an LLM-powered adversary, LLM honeypots offer potential for long-term occupation and denial-of-wallet (DOW) attacks.  
Hence, we argue that the field should optimize systems towards this attacker, and measure progress against this more appropriate class of threats.

Finally, we find that while automated threat intelligence generation is now possible, it remains in a nascent, proof-of-concept stage, primarily blocked by the lack of large-scale, labeled training data.

Based on these findings, we propose a forward-looking research roadmap centered on the development of autonomous and self-improving deception systems.
Architecturally, we highlight the need for open-source, modular frameworks and research into efficient, locally-hosted (hence, privacy-preserving) models using fine-tuning for domain accuracy. 
We advocate for the creation of a dynamic adversarial research ecosystem to enable meaningful evaluation and iterative optimization of systems against the ideal target: sophisticated, automated attackers.
Finally, we outline a path toward greater autonomy through the implementation of feedback loops, enabling systems to automatically generate intelligence for security operations centers (SOCs) and reconfigure themselves to adapt to the evolving threat landscape. 

\section{Prerequisites: Security, Cybersecurity, Honeypots, \& LLMs}\label{sec:background}
The goal of security---viewed broadly, be it for a bike, a home, or a network of computing assets---is to control access, ensure availability, and prevent harm.
The practice of security is to create asymmetry, essentially making a breach prohibitively difficult.
Implicit in this definition is the realization that no system is 100\% secure; for example, in a building, adding doors, locks, thicker walls, and alarm systems cannot guarantee the security of its contents, but may make a successful attack prohibitively difficult or costly.

\subsection{Network Cybersecurity Background}
Cybersecurity is commonly defined as ensuring the confidentiality, integrity, and availability of data and networked computing assets.
The practice of cybersecurity begins with strategically designing a network with a good security posture (e.g., protecting important assets via layers of defense with increasing restrictions, employing the principle of least privilege).
This is complemented by a ``defense-in-depth" strategy, which uses many heterogeneous tools for prevention (e.g., firewalls, anti-malware), detection (e.g., signature-based or anomaly-based detectors), and management (e.g., vulnerability scanning)~\cite{bridges2023soar}.
The overall effect is that well-defended networks enjoy protection against most previously seen attack patterns.
This provides asymmetry, as attackers who use known methods have a decreased probability of success and are forced to develop or acquire new attack patterns, which is presumably more costly.

Note that in network audit logs, it is often very difficult and resource-intensive to distinguish malicious from benign activity, except when known attack patterns are flagged by automated detectors.
This is partly because widespread alert and logging capabilities produce an enormous volume and variety of data that obscures the activity of a sophisticated attacker.
Individual log entries often lack sufficient information, so building a picture of network activity from heterogeneous logs requires a combination of domain expertise and network-specific knowledge---often a slow, labor-intensive process~\cite{bridges2023soar, botta2007towards, bridges2018information,  de2011information, werlinger2009integrated, werlinger2010preparation, goodall2004work}.

Automation has aided both attackers and defenders.
The ease of access to attacking capabilities and the use of automated attacks have increased the scale of threats~\cite{Spitzner2002, valeros2023attacker}, which in turn decreases the cost for the attacker.
On the other hand, automated prevention techniques remain effective against these known, high-volume attacks.
This landscape presents two clear questions for defenders: ``How can we continually learn new attack patterns?'' and ``How can we deter attackers from targeting our network?''

\subsubsection{Cyber Attack Modeling Frameworks}
\label{sec:attack-frameworks}
Frameworks for understanding cyber attacks provide a common language to facilitate attack identification, detection, and threat intelligence sharing.
The term TTPs (Tactics, Techniques, and Procedures) is common in attack analysis.
In this context, \textit{tactics} are high-level adversarial objectives (e.g., Reconnaissance).
\textit{Techniques} are the methods used to achieve a tactic (e.g., port scanning).
Finally, \textit{procedures} are the specific commands or code used to implement a technique (e.g., a specific \texttt{nmap} command).

The Lockheed Martin Cyber Kill Chain~\cite{Hutchins2011CyberKillChain} was influential in framing attacks as logical sequences of tactics.
More commonly used today is the MITRE ATT\&CK (Adversarial Tactics, Techniques, and Common Knowledge) framework~\cite{mitreattack}.
Released in 2015, MITRE ATT\&CK is a globally accessible knowledge base detailing adversary behavior organized into 14 tactics, each with numerous techniques and documented procedures.
Unlike the Kill Chain's linear model, MITRE ATT\&CK provides a broad matrix of TTPs based on real-world observations.
This organized language allows for the labeling and mapping of attack patterns, and the infrastructure to document and share them assists with attribution and cyber defense.

\subsection{Honeypot Background}
Honeypots are decoy computing systems designed to enhance security by gathering threat intelligence and occupying attacker resources.
Consequently, honeypots have been a defensive tactic in cybersecurity since at least the 1980s~\cite{stoll2024cuckoo, cheswick1992evening}.
By convincing unwitting attackers to interact with them, honeypot systems allow for the direct observation of attackers' actions, which in turn informs defensive mechanisms.
This obviates the tedious process of finding malicious activity in a sea of benign system logs.
Furthermore, honeypots are generally isolated environments designed so they can be attacked with no risk to production network assets.

The existence of honeypots also serves to slow cyber attackers.
Novel attack patterns are presumably more expensive to develop than reusing existing tools.
The potential presence of honeypots incentivizes an attacker to ensure they are not interacting with a decoy before using a valuable or secret technique.
Notably, scripts to identify common honeypots exist~\cite{UnaPibaGeek_honeypots_detection}, so even automated attacks may avoid uncamouflaged decoys.
It is believed that attackers will not only seek to detect honeypots but will also attempt cheaper, known attacks before risking a more expensive, novel one~\cite{blackhat_talk_2015}.
All of this slows the attacker, benefiting the defender.

\subsubsection{Honeypot Goals \& Levels of Simulation}
Honeypots are designed to (1) occupy attacker resources and (2) accelerate threat intelligence (e.g., learn TTPs or reduce the time to detect a breach).

If an attacker is interacting with a honeypot, they are, for that time, occupied and not attacking a real host.
This defensive tactic can be the sole goal of a honeypot.
For example, CloudFlare uses generative AI to create decoy web networks to occupy malicious web-scraping bots~\cite{tatoris2025}.
Similarly, researchers have used AI to create fake Active Directory users~\cite{lukas2021deep} or to automatically engage with email and phone scammers to waste their time~\cite{cambiaso2023scamming, AIDaisyGranny}.

In addition to occupying attackers, honeypots are also used to lower the time-to-detection for a cyber breach.
Often called ``canaries,'' these honeypots are designed to entice attacker interaction to discover an active breach faster, thereby preventing further damage~\cite{Spitzner2002, dagon2004honeystat}.

Honeypots are also used to gain threat intelligence by directly observing attacker TTPs.
Examples include honeypots that capture malware~\cite{holbel2024utilizing}, distributed networks that gain a panoramic view of attack trends~\cite{outpost24_threat_landscape_2023}, and recent designs intended to identify and counter LLM-powered offensive agents~\cite{heckel2024countering, reworr2025llmagenthoneypotmonitoring}.

With these goals in mind, honeypots mimic servers with varying degrees of realism and risk.
They are conceptualized on a spectrum from low-interaction to high-interaction.
Low-interaction honeypots are almost fully simulated with limited capabilities.
Because they have little real connectivity, they pose minimal risk to the defender's network, but they lack the realism to entice sophisticated attack behavior.
Medium-interaction honeypots have some real functionality but are restricted to limit risk.
High-interaction honeypots are generally real systems, which require more resources to deploy and entail greater risk.
Many templates for these honeypot types are open-sourced~\cite{the_honeynet_project}.
In general, lower interaction means faster deployment and less risk, but at the cost of being less convincing.
Experts have stated that low-interaction honeypots will likely never convince a skilled human attacker~\cite{breaking_honeypots_blackhat_2015}, and our survey (Section \ref{sec:honeypot-detection}) shows there are many ways to detect a simulated system.
The vision driving the integration of LLMs into honeypot systems is to achieve the fidelity of a high-interaction honeypot with the speed, ease, and low risk of a lower-interaction system.

\subsubsection{Honeypot Necessities}
Overall, we posit two observations.
Firstly, for a honeypot to succeed, it must be sufficiently convincing to elicit attacks in the wild.
Secondly, if one seeks to gain threat intelligence, the data produced must be efficiently converted into actionable information, meaning the cost to the defender must be less than the value of the intelligence gained.

\subsection{LLM Background}
Large Language Models (LLMs) are generative machine learning models that produce text and possess an extremely broad knowledge base.
They use the transformer architecture~\cite{vaswani2017attention}, which utilizes attention mechanisms to effectively capture long-range dependencies between tokens.
Tokens can be understood as generalizations of words or sub-words.

LLMs typically undergo at least two primary training phases.
The initial phase, pre-training, involves training the model on next-token prediction using vast amounts of text data.
The second phase, fine-tuning, adapts the model using curated input-output text pairs, where inputs might be questions or instructions, and outputs are the desired responses.
This two-stage process enables LLMs to perform a wide array of tasks, such as text generation, summarization, and code writing.

Prominent examples include OpenAI's GPT series~\cite{openai_models}, Meta's Llama family~\cite{meta_llama}, and Google's Gemini models~\cite{google_gemini}.
Furthermore, specialized LLMs are being developed for security applications, such as the offensive security-focused WhiteRabbitNeo \cite{whiterabbitneo} and research into cybersecurity-tuned LLMs \cite{jaffal2025ai}.
The rapid expansion of LLM research is documented in various surveys~\cite{fan2023bibliometric, qiao2022reasoning, zhao2023survey}.
LLMs are often augmented with tools, such as a web search capability or a Python interpreter, that allow for greater functionality.
Of particular use are retrieval-augmented generation (RAG) systems, in which an LLM queries a vector store database to gain extended context.

The ability of LLMs to rapidly generate content that deceives humans has ignited exploration of their use in cybersecurity, both for offensive purposes~\cite{gupta2023chatgpt, Spitzner2002, valeros2023attacker, heckel2024countering, reworr2025llmagenthoneypotmonitoring} and, as discussed in Section \ref{sec:llm-honeypot}, for defensive ones.
A clear dichotomy exists in use between large proprietary models accessible via API and open-weight LLMs that can be run locally, but are generally less capable.

\section{Related Fields}
\label{sec:surveys}
In this section we survey three areas, namely, 
 honeypot fingerprinting methods,  
 LLM shell honeypots, 
 and honeypot data analysis methods.
 In each, we organize the works into either categories or developing trends to facilitate a clearer understanding.

    \subsection{Honeypot Detection Vectors}\label{sec:honeypot-detection}
Because effective camouflage is necessary for honeypot success, we provide a representative (but not exhaustive) survey of honeypot detection (or fingerprinting) methods. 
The goal is to identify the fundamental vectors an adversary can exploit to expose a decoy.
These vectors form the basis of our systematization and provide a roadmap for where LLMs can offer the most significant improvements.
We found four categories of detection vectors or tactics that persist over time (although the specific tactics vary widely). Each is described in a subsection below and a summary is given in 
Table \ref{tab:honeypot-vectors-llm}, including LLM simulation's contributions. 

\begin{table*}[ht]
\centering
\caption{Taxonomy of Honeypot Detection Vectors and LLM Simulation's Impact}
\label{tab:honeypot-vectors-llm}
\small
\renewcommand{\tabularxcolumn}[1]{m{#1}}
\begin{tabularx}{\textwidth}{>{\raggedright\arraybackslash}m{1.45cm} *{4}{>{\raggedright\arraybackslash}X}}
\toprule
 & \textbf{Contents \& Posture} & \textbf{Outputs \& Behavior} & \textbf{Functional Limits} & \textbf{Feature Synthesis} \\ 
 \midrule
\textbf{Description} & Inconsistencies in claimed identity vs. observable properties & Realism of response content and associated metadata. & Absence of underlying system logic or side-effects. & Multi-feature analysis using machine learning. \\ 
\rowcolor{lightgray}
\textbf{Examples} & Default banners; unrealistic services or file system contents & Static \texttt{htop} output; response timing anomalies & \texttt{vim} interaction fails; \texttt{wget} egress fails; binary execution fails & Classifiers, later LLMs, with multiple vectors\\ 
\textbf{LLM Aid} & Generates diverse, believable, context-aware  content& Provides context-aware command responses with stochasticity & Simulates logical ``success'' messages and UI feedback & Increases feature-space complexity, breaking static signatures \\ 
\rowcolor{lightgray}
\textbf{LLM Fail} & Cannot obscure open ports or network stack (TCP/IP) signatures & Cannot reproduce temporal fidelity or real-time dynamics & Cannot generate actual traffic, file creation, or kernel execution & Combined vectors increase detection confidence \\ 
\bottomrule
\end{tabularx}
\end{table*}

\subsubsection{Contents \& Network Posture}
Inconsistencies in the honeypot's claimed identity versus its observable properties are a common giveaway.
This includes unrealistic ports, services, and banner messages.
Similarly, the presence of unrealistic or default data within the system is another key indicator.
This ranges from default file system layouts and user accounts in open-source honeypots to a general lack of plausible, lived-in content, such as applications and file contents.

Early detection methods focused on probing for environmental and network inconsistencies.
The early work of Provos \cite{provos2004virtual} created a honeypot network named \texttt{Honeyd}.
Provos took care to camouflage the system’s fingerprint from scanning tools like NMAP, as well as adjusting outgoing packets (e.g., ISN, IP, and TCP characteristics) to maintain consistency.
Subsequent research quickly found new inconsistencies with honeypots.
Holz \& Raynal showed that simple Bash commands or hardware details (e.g., MAC addresses) could reveal a simulated environment~\cite{holz2005detecting}.
Defibaugh-Chavez et al. \cite{defibaugh2006network} illustrate many methods of honeypot detection, namely, exercising many different services (identifying ones that should exist but either do not or are unconvincing) and examining packet contents.
These red flags have been reported for almost 20 years.

Recent works demonstrate that honeypots often expose an unrealistic network posture, such as running too many disparate services (e.g., multiple databases purportedly running on one IoT device), having mismatched service banners, or possessing insufficient resources (memory, disk space) for the machine they claim to be~\cite{wenda2011honeypot, censys_honeypots_2023}.

A particularly effective and simple vector is the use of default configurations in open-source honeypots, which creates thousands of easily identifiable decoys in the wild~\cite{Morishita2019DetectMe, srinivasa2021gottacatchemall, breaking_honeypots_blackhat_2015}.
For example, Morishita et al. \cite{Morishita2019DetectMe} crafted 20 signatures to identify default responses from open-source honeypots and discovered over 19,000 honeypots in the wild.
Cabral et al. \cite{cabral2019review, cabral2021advanced} have shown empirically that mitigating these static configuration issues by changing banners, file systems, and network services significantly improves evasion.
Their study investigates Cowrie \cite{Oosterhof2024Cowrie}, a popular open-source, medium-interaction SSH and Telnet honeypot, focusing on how configurations can lead to detection.
The follow-on work provides tips on how to configure Cowrie to evade detection, such as changing banners, file system contents, network interactions, and the reported software and OS descriptions.
Guan et al. \cite{guan2024honeyllm} identify four types of problems with LLM honeypot responses: the LLM does not know the input command; the LLM responds with an incorrect format; the LLM provides an unbelievable or incorrect simulated response; and the LLM responds in a manner inconsistent with history.
Vetterl \& Clayton \cite{Vetterl2018} show that open-source honeypots' implementations of SSH, TCP, and HTTP protocols can be detected through protocol-specific probing and analysis.
Their detection method shows that camouflaging honeypot protocols is necessary to avoid known fingerprinting methods.

\subsubsection{Outputs \& Behavior}
The second major category of fingerprinting stems from outputs that are themselves incorrect or unrealistic, or metadata associated with the output that is unbelievable.
This includes unbelievable TCP/IP characteristics, detectable time delays in communications, or other anomalies, e.g., the \texttt{top} command producing a static, nonuniformly spaced table instead of a continuously updated one.

Attackers can identify honeypots not by what they are, but by how they behave.
Defibaugh-Chavez et al. \cite{defibaugh2006network} also noted that simulated systems often respond much slower to simple ICMP (ping) requests.
This timing-based fingerprinting was later formalized to identify Honeyd deployments with high accuracy, as their packet link latencies could differ by an order of magnitude from real systems~\cite{fu2006recognizing}.
This concept was further extended by applying machine learning classifiers to timing statistics from ICMP and TCP/IP to automatically flag decoys~\cite{mukkamala2007detection}.

Moving to LLM-based honeypots, timing remains a critical and unresolved challenge~\cite{sladic2024llm, ragsdale2023designing, fan2024honeyllm}.
Further, Wang et al. \cite{Wang2024} find that ``Common impediments such as unsupported tools, insufficient privileges, or the absence of targeted information typically result in the termination of the attack engagements. Therefore, a higher command execution success rate is critical.''
This is echoed by Ragsdale \& Boppana \cite{ragsdale2023designing}, who cite interactive commands like \texttt{Vim} as a pitfall for honeypots.
Furthermore, Ragsdale \& Boppana, Wang et al., and Guan et al. \cite{guan2024honeyllm} identify potential vulnerabilities for LLM-based honeypots, particularly denial-of-service via input flooding and prompt injection attacks.

\subsubsection{Functional Limitations}
A third, more fundamental vector is the exploitation of a honeypot's inherent functional limitations.
This includes simple user interaction problems, such as failed shell tab-completion, and extends to more substantial limitations, like restricted outbound network traffic.

The first documented detection technique of this kind was observed in the wild by Krawetz \cite{krawetz2004anti} in 2004 while studying email spammers.
The spammers required detection of proxy servers that they used to hide their IP.
Their technology, called Honeypot Hunter, used a suspected proxy server to connect back to the attacker's own host to see if the response was realistic; if the proxy server claimed a successful connection, but none was observed, then the proxy was identified as a honeypot.
Krawetz notes that the Honeypot Hunter detection methodology is itself detectable and predicts a cat-and-mouse style progression of detection methodologies.

Such avenues for detection persist for LLM-powered honeypots.
Ragsdale \& Boppana \cite{ragsdale2023designing} identify external connections as a detection vulnerability, as LLMs can simulate successful external communications but have no means of actually performing the connection.
``Such scenarios include starting a session with a C\&C [command and control] server, downloading malware, or exfiltrating data. Since no traffic is generated, the attacker can determine that their commands are not being executed and that they are in a honeypot.''
This principle of ``egress control'' detection has been leveraged in various ways.
Botnets can identify honeypots by checking if a compromised host can be used to launch further attacks~\cite{wang2010honeypot}, and firewalls designed to neuter a honeypot can be detected when they block outbound traffic that a real system would permit~\cite{wenda2011honeypot}.

\subsubsection{Synthesizing Multiple Features}
Finally, modern approaches synthesize all these vectors using machine learning.
Huang et al. \cite{huang2019automatic} gathered features of a suspect honeypot and used SVM, $k$-NN, and Naive Bayes classifiers to identify honeypots.
The features used were based on protocols (e.g., whether services exist), network flow (e.g., average TTL), and system characteristics (e.g., ICMP response times).
To obtain labeled data, the authors used online internet scanning and honeypot identification sites and built features from real servers.
The approach exhibited high detection accuracy.
Srinivasa et al. \cite{srinivasa2021gottacatchemall} searched for honeypot identifiers from previous works and signatures from Morishita et al. \cite{Morishita2019DetectMe} to identify open-source honeypots in the wild, finding over 21,000.
The authors emphasize that these techniques target low- to medium-interaction honeypots only.

Ilg et al. \cite{ilg2025beekeeper} present \texttt{Beekeeper}, an LLM-driven system to analyze honeypot realism.
Notably, they found that attempted downloads and use of files often gives away honeypots, as the decoys simulate but do not have actual functionality.
This work shows not only that LLMs can be used to identify honeypot weaknesses, but that LLMs (and by extension, LLM-powered attackers) can fingerprint honeypots automatically.


\begin{table*}[ht!]
\centering
\caption{LLM-Powered Shell Honeypot Works}
\label{tab:llm-honeypot}
\begin{tabular}{@{} p{2cm} p{1.8cm} p{5.2cm} p{2cm} p{3.3cm} @{}}
\toprule
\textbf{Paper} & \textbf{Type} & \textbf{Core Contribution} & \textbf{LLMs Tested} & \textbf{Evaluation Technique} \\ \midrule

McKee \& Noever \cite{McKee2023} (2023) & General (OS, App, Network) & The first conceptual proposal for using LLMs for a variety of honeypot-related tasks, demonstrated via prompting. & ChatGPT (Dec 2022) & Conceptual Prompting (In Lab) \\

\rowcolor{lightgray}
Ragsdale \& Boppana \cite{ragsdale2023designing} (2023) & Shell (SSH, Telnet) & Proposed novel methods for efficiency, including intelligent context pruning and caching deterministic responses for the shell. & GPT-3.5 & Statistical (Levenshtein, In Lab); Script Replay (Campaign Length, In Lab)\\

Sladić et al. \cite{sladic2024llm} (2024) & Shell (SSH) & Presented the first LLM-integrated shell honeypot (\texttt{ShelLM}). Identified foundational challenges like latency and context limits. Released the Prague Dataset. & GPT-3.5 & Human Study (12 Experts, In Lab) \\

\rowcolor{lightgray}
Wang et al. \cite{Wang2024} (2024) & Shell (SSH, Telnet) & Developed a sophisticated Prompt Manager to track system state in a shell honeypot. Introduced a hybrid architecture and a suite of new metrics. & GPT-3.5, 4 & Script Replay (New Metrics, In Lab)
; Comparative Engagement (In the Wild) \\

Guan et al. \cite{guan2024honeyllm} (2024) & Shell (SSH, Telnet) & Systematized the filter/router architecture for shell honeypots to handle scanners and manage costs. Provided strong evidence for chain-of-thought prompting. & GPT-3.5, 4o, Claude-2, 3 Haiku, 3 Opus, Llama 2 (70B) & Script Replay (Fidelity, In Lab); Statistical (Session Length, In the Wild) \\

\rowcolor{lightgray}
Weber et al. \cite{weber2024don} (2024) & Shell (SSH) & Conducted a critical analysis of shell fidelity limitations, identifying specific failure modes. Found only 52\% of responses were convincing. & GPT-3.5 & Human Study (5 Experts, In Lab); Statistical (SBERT Similarity, In Lab) \\

Otal \& Canbaz \cite{otal2024llm} (2024) & Shell (SSH) & Investigated fine-tuning an open-weights LLM on shell data to improve the quality of simulated responses. & Llama3-8B & Statistical (Similarity, In Lab) \\

\rowcolor{lightgray}
Fan et al. \cite{fan2024honeyllm} (2024) & Shell (SSH) & Proposed seven fidelity metrics for shell simulation. Evaluated multiple commercial LLMs, concluding GPT-4o was superior for the task. & GPT-4, 4o, Gemini Pro 1.5, Claude 3 Opus, Mistral 7B & Statistical (Session Time, In the Wild) \\

Johnson et al. \cite{johnson2024modular} (2024) & Shell (SSH) & Presents a modular architecture (\texttt{LIMBOSH}) with prompt injection mitigation and a separate LLM context for output fidelity. & GPT-4o & Human Study (4 Experts, vs. real server, In Lab) \\

\rowcolor{lightgray}
Christli et al. \cite{christli2024ai} (2024) & General & Evaluates Llama 3's shell simulation accuracy against a real system & Llama 3 & Statistical (In Lab) \\

Gizzarelli \cite{Gizzarelli2024} (2024) & Shell (SSH, MySQL)  & {Introduced SYNAPSE with automated log-to-MITRE ATT\&CK mapping. Performed a comparative human study where SYNAPSE was perceived as more realistic than its static equivalent.} & GPT-3.5, 4, 4o, Gemini & Human Study (vs. static, In Lab); Comparative (In the Wild)\\ 

\rowcolor{lightgray}
Badran \& Niazi \cite{badran2025towards} (2025) & Shell (HTTP) & Used a prompted LLM to both label HTTP requests by attack type and generate believable server responses. & GPT-4o & Human Study (10 Annotators, In Lab) \\

Malhotra et al. \cite{malhotra2025llmhoney} (2025) & Shell (SSH) & Hybrid architecture with state-aware LLM to balance latency and fidelty used to evaluate many open-weight LLMs for shell simulation. & 13 open-weight LLMs, see text & Statistical (Accuracy, latency, hallucinations, In Lab) \\

\rowcolor{lightgray}
Jimenez et al. \cite{jimenez2025design} (2025) & LDAP& Design and implement LDAP honeypot, curate LDAP dataset for fine-tuning, and build evaluation methodology. & Llama 3 (8B)& Statistical,  (syntax, structural and content accuracy, completeness metrics; In Lab)\\

Sladić et al. \cite{sladic2025vellmes} (2025) & Shell (SSH, MySQL, POP3, HTTP) & Architecture identifies protocol, and uses different protocol-specific prompts to LLM &  GPT-3.5, 4 & Statistical (unit-tests), human study (89 participants), real-world deployment (In Lab \& In the Wild) \\

\rowcolor{lightgray}
Safargalieva et al. \cite{safargalieva2025ohra} & Shell (SSH, Telnet, HTTP, FTP, SMTP, IPP, SNMP) & Design and implement LLM-powered honeypot for seven protocols, evaluated for accuracy, deceptiveness, and latency. & 10 models, see text & Statistical and qualitative tests for latency, accuracy (In Lab \& In the Wild) 
\\

\bottomrule
\end{tabular}
\end{table*}
    \subsection{Honeypots Using LLMs}
\label{sec:llm-honeypot}
This section surveys papers proposing methods to advance cybersecurity honeypot technologies using LLMs.
To the best of our knowledge, this is a comprehensive survey of research that integrates LLMs into network security honeypots.
We identified papers in this field using a Google Scholar search for \texttt{"LLM" AND "Honeypot"} through October 2025.
We structure our review thematically to systematize the field's evolution, showing how foundational concepts matured into fidelity evaluations, architectural refinements, and in-the-wild deployments.
Table \ref{tab:llm-honeypot} itemizes and summarizes every work in the area.

\subsubsection{Foundational Concepts \& Challenges}
The concept of using LLMs to create dynamic honeypots emerged in late 2022.
McKee \& Noever \cite{McKee2023} were likely the first to propose using chatbots (LLMs) in honeypot environments.
They itemized ten honeypot-related tasks and demonstrated (in December 2022) ChatGPT's ability to perform them with straightforward prompting, thereby illustrating the diverse potential capabilities LLMs offer for honeypots.

Following this proposal, the first implementation was presented by Sladić et al. \cite{sladic2024llm}, who integrated an LLM into a Linux shell honeypot.
Their system, \texttt{ShelLM}, uses OpenAI's GPT-3.5-turbo-16k API to simulate shell responses, passing each command along with the complete command history to the LLM.
In their evaluation, a user study showed that human experts often could not identify the simulated responses.
However, in building the first system, they also discovered the foundational challenges that would -define subsequent research: LLM response latency, potentially insufficient context window lengths, and the stochastic nature of LLM outputs, which could lead to inaccuracies.
Notably, the authors also publicly released the Prague Dataset, a small dataset of real shell sessions.

\subsubsection{LLM Shell Simulation Unit Testing}
A lineage of works focused primarily on ``unit-tests'' (a term coined by Sladić et al. \cite{sladic2025vellmes}) of LLMs' abilities to accurately simulate real computer interaction.
Many different metrics and experimental methods were developed.

To improve the quality of simulated responses, Otal \& Canbaz \cite{otal2024llm} investigated fine-tuning Llama3-8B on shell command data from Cowrie and empirically verified that the fine-tuning resulted in outputs more similar to Cowrie's.
Weber et al. \cite{weber2024don} conducted an experiment where five Computer Science graduate students graded GPT-3.5's responses for believability (binary) on over 1,400 unique request-response pairs, encompassing 230 different base commands.
They concluded that only 52\% of the generated responses were convincing.
They find that long and compound commands are problematic for LLMs and discuss command features with a high likelihood of producing unbelievable outputs.
Their results indicate that the cosine similarity of SBERT text embeddings between simulated and real responses strongly predicts believability, suggesting a path beyond simple string metrics.
Malhotra \cite{malhotra2025llmhoney} introduces \texttt{LLMHoney}, another shell honeypot system, and instantiates it for comparative testing with 13 open-weight LLMs.
Using 138 ``representative shell commands,'' Malhotra evaluates each instantiation using numerous metrics to gauge fidelity, latency, hallucination rate, and memory footprint.
He found that latency and memory increase with model size, but fidelity suffers with small models.
Christli et al. \cite{christli2024ai} also used Llama 3 to simulate shell responses, evaluating response fidelity against a real system with similarity metrics.

Fan et al. \cite{fan2024honeyllm} propose seven metrics for evaluating the fidelity of simulated shell responses.
Leveraging these, they tested many powerful LLMs requiring API calls, evaluating them for the fidelity, latency, throughput, and cost of their shell response emulation.
Based on their findings, they concluded GPT-4o is best overall and used a Raspberry Pi to implement a simple architecture for filtering problematic inputs, passing only acceptable inputs to the LLM API for server emulation.
They observed a higher frequency of longer sessions (in terms of clock time) for GPT-4o over the non-LLM honeypots Cowrie and Amun \cite{gobel2009amun}.

\subsubsection{Architectures Advancement \& Real-World Deployments}
In parallel, meaningful architectural advancements for honeypot security, practicality, and cost were also pioneered.
The foundational challenges of latency and cost led to a critical architectural evolution: the hybrid or filtered honeypot, which leverages a component to manage the volume and content of requests sent to the LLM.
This design seeks to use the expensive LLM only when necessary, protect against Denial of Service/Wallet (DoS/DoW) attacks, reduce accumulating context throughout an attack session, and enhance fidelity.
Many of these works also contributed to testing shell simulation fidelity and architectural advancement began evaluating the whole system  in-the-wild.

Ragsdale \& Boppana \cite{ragsdale2023designing}, also a pioneering work in the area, investigated GPT models for simulating shell responses, observing limitations similar to those reported by Sladić et al. \cite{sladic2024llm}.
They provide novel methods for handling the volume and quality of what is sent to the LLM by intelligently trimming previous commands that do not alter context.
This approach demonstrates significant gains in token efficiency.
In the same vein, they propose caching deterministic responses to reduce latency and load on the LLM.
To measure accuracy and deceptiveness, they measured per-command fidelity using Levenshtein distance and found the LLM-powered honeypot to be slightly better than Cowrie.
Next, they ran sequences of attack commands comprising an attack campaign against the honeypots to quantify how much of the campaign could be completed before a command fails.
The LLM-powered honeypots exhibited longer sessions, with Cowrie often failing in the first few commands.
This corresponds to Wang et al.'s \cite{Wang2024} observation that the inability to provide a valid command response is a major obstacle for low-interaction honeypots.

Wang et al. \cite{Wang2024} introduced \texttt{HoneyGPT}, a system designed to address the honeypot ``trilemma'' of balancing flexibility, interaction, and deception.
Its primary innovation is a sophisticated Prompt Manager that orchestrates LLM interaction.
To manage costs, the manager prunes the interaction history to prevent exceeding context limits and routes simple commands to traditional emulators or a cache.
This is an important advancement in line with developments by Ragsdale \& Boppana \cite{ragsdale2023designing} and Guan et al. \cite{guan2024honeyllm}, as LLM-powered honeypots introduce a denial-of-service (DoS) vulnerability from overloading the LLM.
For honeypot evaluation, the authors created metrics to statistically evaluate honeypots, including methods for quantifying command execution success and logic, session length, and attacker response likelihood.
A four-week in-the-wild deployment comparison shows that HoneyGPT achieves a deeper interaction with the attacker than Cowrie and elicits six ATT\&CK techniques that Cowrie did not.

Building on a similar hybrid concept, Guan et al. \cite{guan2024honeyllm} also considered LLMs simulating shell responses in a honeypot.
Like Wang et al., Guan et al. instantiated a honeypot system with a filtering/routing step to prevent scanning activity from engaging the LLM.
The filter leverages signatures to identify scanning scripts.
Notably, Guan et al. provided statistics showing that a honeypot's typical query rate exceeds API limits, proving that this DoS vulnerability is naturally exploited by the high rate of attacks.
Their filtering logic keeps rates well within these limits.
Guan et al. focus in part on prompting and provide guidance and tests for it.
Finally, they provided statistics from ``in-vitro'' (laboratory) tests and an in-the-wild deployment using many different LLMs alongside a Cowrie-only honeypot.
For laboratory tests, they constructed 50 attack sessions and tested each LLM-honeypot with a ``vanilla'' versus a chain-of-thought (CoT) prompt, identifying the session length until an unbelievable output occurs.
Across the board, chain-of-thought prompting provides a large boost.
From a three-week internet deployment, they found that for effective attacks (those not deemed to be scanning), the average session length was 2.96 commands with Cowrie, which was statistically significantly smaller than an average of 5.83 commands from the best LLM-powered version.

Johnson et al. \cite{johnson2024modular} present \texttt{LIMBOSH}, a modular software architecture for an LLM shell.
It incorporates prompt injection mitigation with input transformations and a random forest prompt injection classifier.
Additionally, an output component employs a separate LLM context to ensure fidelity.
A single-blind, within-subjects user study involved four cybersecurity professionals red-teaming a real Ubuntu server and \texttt{LIMBOSH} with GPT-4o, without prior knowledge that either was a honeypot.
Likert-scale questions about each were compared, showing the LLM honeypot exhibited perceived realism.

\subsubsection{Expanding the Attack Surface}
With the core concepts established in shell-based honeypots, researchers began applying the same principles to other protocols and services.

Gizzarelli \cite{Gizzarelli2024} introduced \texttt{SYNAPSE}, a dynamic honeypot that simulates both a Linux terminal and a MySQL server using GPT-4o.
The system was evaluated through a unique three-part study: a ten-day in-the-wild deployment, an AI vs. AI attack scenario, and a comparative human study.
In the human study, 18 participants interacted with both \texttt{SYNAPSE} and its traditional static equivalent, with nearly 80\% perceiving the AI-powered version as more realistic.
Hu et al. \cite{hu2024my} used an LLM in an SQL honeypot to generate responses to attacker queries.
Their system, \texttt{MySQL-Pot}, exhibited longer interaction session lengths compared to an analogous non-LLM powered system.
Sezgig \& Boyacı \cite{sezgig-boyacı} created a web API honeypot employing a Retrieval-Augmented Generation (RAG) system.
This system queries a database of API requests and responses to generate realistic, automated replies.
Badran \& Niazi \cite{badran2025towards} tested GPT-4o in responding to HTTP requests, with a system prompt that directs the LLM to first label the request as an attack type and then respond as a web server.
In a human evaluation, they found 87\% accuracy on attack labels and that 80\% of responses were considered convincing.

Jimenez et al. \cite{jimenez2025design} built an LDAP honeypot based on Llama 3 (8B), and
describe the process of implementing a honeypot on a lesser-known protocol, including dataset curation, fine-tuning an open-weight LLM, architecture, and customized evaluations.
They find that the base model is poor at LDAP communications, but fine-tuning with a well-curated dataset provides dramatic gains.

Sladić et al. \cite{sladic2025vellmes} extended their SSH honeypot \texttt{shelLM} to the \texttt{VelLMes} system, now including support for MySQL, POP3, and HTTP protocols.
Each protocol is implemented by prompting an LLM using chain-of-thought techniques.
Architecturally, the system identifies the protocol from the input and sends the correct prompt and history to the LLM.
Evaluations were performed to test simulation accuracy via statistical unit tests, deceptiveness via an 89-person user study, and efficacy in a real-world deployment.
GPT-3.5 and GPT-4 were evaluated.

Safargalieva et al. \cite{safargalieva2025ohra} present \texttt{OHRA}, a web-facing honeypot that fully supports SSH, Telnet, and HTTP, and partially supports several other protocols.
Ten LLMs were tested, and the authors found GPT-4o-mini to be best overall.
The authors discuss simulation strengths and weaknesses per protocol and interestingly found that latency could be unrealistically slow at times and unusually fast at others.

\subsubsection{Open-Source LLM Honeypots}\label{sec:open-source-honeypots}
The concepts from this research are being implemented in publicly available tools.
Beelzebub is a honeypot with LLM simulations that supports TCP and HTTP, allowing a wide range of services to be emulated; in particular, the developers claim ``full support for SSH''~\cite{BeelzebubHoneypot}.
Galah is a honeypot project enabling API connections to various LLMs to ``dynamically craft relevant responses—including HTTP headers and body content—to any HTTP request''~\cite{karimi2024galah}.
The \texttt{T-Pot} platform facilitates the deployment of numerous open-source honeypots and includes a dedicated section for LLM-based honeypots, supporting Beelzebub and Galah deployments with Ollama LLMs~\cite{TelekomSecurityTPotLLMHoneypot}.
The platform's maintainers note: ``We think LLM-Based Honeypots mark the beginning of a game change for the deception / honeypot field.''
    \subsection{Honeypot Log Analysis}\label{sec:log-analysis}
While many studies report on attack statistics from honeypot deployments~\cite{Kemppainen2018, kelly2021comparative}, this survey concentrates on research aimed at automating the transformation of raw honeypot data into actionable threat intelligence.
Our systematization reveals a clear evolution in this research, progressing through four  phases of  maturity.
Importantly, the recent automated methods are promising as they enable previously unprecedented attack labeling via fine-tuning or few-shot learning with LLMs.

\subsubsection{Data Reduction \& Anomaly Detection}
The earliest challenge in honeypot analysis was managing the sheer volume of log data.
The goal of this initial phase of research was not to fully understand attacks, but simply to reduce the noise and find interesting sessions worthy of a human analyst's time.
Thonnard \& Dacier~\cite{Thonnard2008} used clustering based on timing characteristics in honeypot data to reduce the large number of individual attacks to a smaller set of attack clusters suitable for manual analysis.
Ghourabi et al.~\cite{Ghourabi2013CharacterizationOA} proposed analysis methods for a web service honeypot, using unsupervised techniques to find a manageable subset of activity.
They trained a support vector regressor to predict message sizes and a classifier to predict message classes, flagging messages with significant deviations for manual inspection.
More recently, Aslan et al.~\cite{Aslan2024Unveiling} used Latent Dirichlet Allocation (LDA) to analyze honeypot logs, learning topics of co-occurring commands to reveal patterns, such as the frequent use of \texttt{wget} when downloading malware.

\subsubsection{Session Classification \& Visualization}
This phase moved beyond just flagging anomalies to trying to categorize and understand them.
The focus shifted to building tools that could either label entire sessions or help a human explore the data more effectively.
Spyros et al.~\cite{spyros2022towards}, for example, used Dionaea honeypot logs to train multiple classifiers (e.g., AdaBoost, Random Forest) to categorize attack activity as high-impact or low-impact based on features from each session.
Others focused on generating outputs for other security tools.
Owezarski~\cite{Owezarski2015ANR} identified anomalies in network flows within honeypot attack data to infer filtering rules that could later be used by network intrusion detection systems.
Another thread of research focused on human-in-the-loop exploration through visualization.
Fraunholz et al.~\cite{fraunholz2017data} proposed a dashboard consisting of multiple visualizations of informative statistics from honeypot data, a theme also explored by others~\cite{ikuomenisan2022systematic, valli2009visualization}.
Mehta et al.~\cite{mehta2021} also used the ELK stack for visualization but additionally employed machine learning to predict file and folder traversal, thereby forecasting an attacker's next steps.

\subsubsection{Automated Mapping to MITRE ATT\&CK}
A major turning point in log analysis was the shift from statistical classification to semantic understanding: translating raw commands into the standardized language of attacker behavior via MITRE ATT\&CK TTPs.
This marks an evolution from processing raw data to generating threat intelligence.

An early example is the XT-Pot framework from Ryandy et al.~\cite{Ryandy2020XTPOTET}, which mapped observed attacker activity to ATT\&CK techniques, presumably using manually defined rules to create ``soft signatures."
Gizzarelli \cite{Gizzarelli2024} introduced an LLM-powered honeypot, \texttt{SYNAPSE}, which includes a machine learning classifier to automatically map attacker activity to the MITRE ATT\&CK framework.
When evaluated against real-world attack data, this mapping extension achieved 75\% precision and 68\% recall.

Boffa et al. shows a clear progression in this area that mirrors the evolution of the natural language modeling community.
They first used Word2Vec embeddings of bash commands to create clusters that were then manually annotated with attacker goals~\cite{boffa2022towards}.
Later, they presented LogPrécis, a fine-tuned BERT model that automatically applies ATT\&CK tactic labels to each command in a honeypot session, achieving over 90\% accuracy~\cite{boffa2024logprecis}.

With the advent of modern LLMs, researchers began using prompting for this labeling task.
Ozkok et al.~\cite{ozkok2024honeypot} developed system prompts for GPT-4 to explain attack consequences and label logs with ATT\&CK techniques, achieving 72.46\% accuracy on the latter task.
Similarly, Badran \& Niazi \cite{badran2025towards} prompted GPT-4o to apply one of five labels to HTTP requests, finding 87\% accuracy.
Lanka et al.~\cite{lanka2024} prompted an LLM to generate plain-text descriptions of TTPs from shell commands, which were then used to build a vector store for attack detection.
However, this approach has limits.
Daniel et al.~\cite{daniel2025labeling}, in a related problem of labeling SNORT rules, found that while LLMs provide explainable and efficient mappings, traditional ML models ``consistently outperform them in accuracy.''
This highlights the significant difficulty of the semantic mapping task, a challenge summarized by Jiang et al.~\cite{jiang2025mitre}: ``Mapping real-world behaviors to ATT\&CK techniques is a resource-intensive and subjective process, often prone to bias... Effective mapping relies heavily on dataset quality and expert involvement ..."

\subsubsection{Operationalizing Intelligence with Agents}
The most recent phase of research seeks to move beyond post-facto log analysis and use the intelligence gathered from honeypots to power real-time detection on production systems.
Lanka et al.~\cite{lanka2024} exemplify this frontier with a system that leverages a shell honeypot with a Retrieval-Augmented Generation (RAG) system.
After preprocessing, attack commands from the honeypot are stored as vectors in a RAG knowledge base.
For real-time detection, commands from a user on a live system are compared against the known malicious commands in the RAG system, with a prompt to GPT-4o crafted for final classification.
This system shows a new potential for using agents with shell honeypots for more sophisticated detection.
This work connects to a parallel, emerging area of using LLM agents to enhance threat intelligence and detection~\cite{Bokkena2024enhancing}, suggesting a promising future in merging these fields.


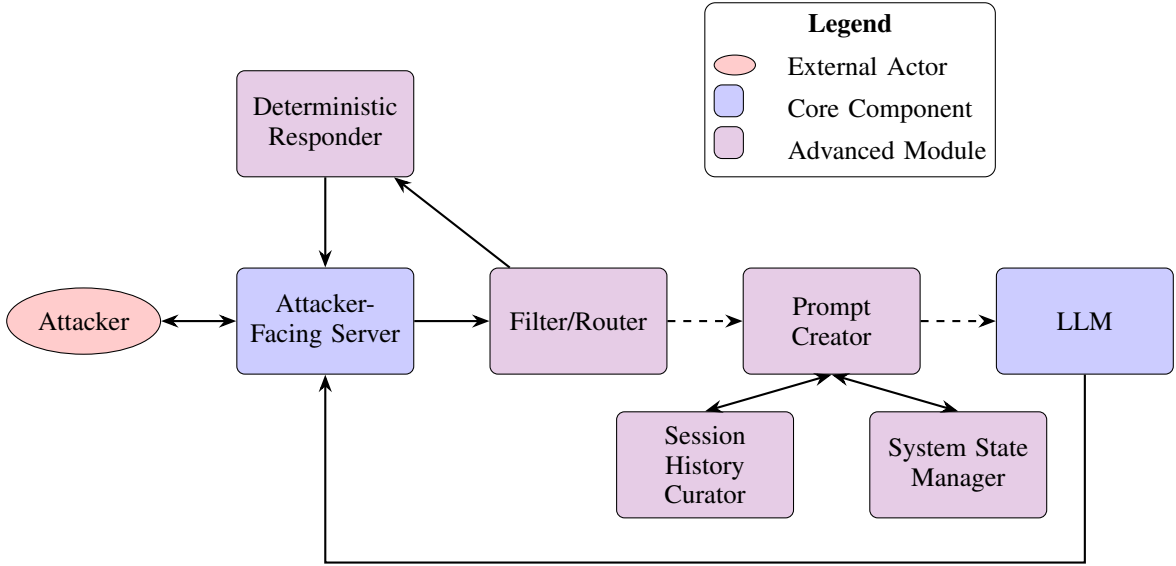
\begin{figure*}[ht]
    \centering
    \begin{tikzpicture}[
        node distance=1.2cm and 1cm,
        actor/.style={ellipse, draw, fill=red!20, minimum height=2.5em, text centered},
        core/.style={rectangle, draw, fill=blue!20, minimum height=4em, text width=6em, text centered, rounded corners=3pt},
        arrow/.style={thick, ->, >=Stealth},
        module/.style={rectangle, draw, fill=violet!20, minimum height=4em, text width=6em, text centered, rounded corners=3pt},
        twowayarrow/.style={thick, <->, >=Stealth}, 
        internal/.style={dashed, thick, ->, >=Stealth} 
    ]

    \node[actor] (attacker) {Attacker};
    \node[core, right=of attacker] (server) {Attacker-Facing Server};
    \node[module, right=of server] (filter) {Filter/Router};
    \node[module,right=of filter] (prompt) {Prompt Creator};
    \node[core, right=of prompt] (llm) {LLM};
    \node[module, above left=of filter] (responder) {Deterministic Responder};
    \node[module, below=of $(filter)!0.5!(prompt)$] (history) {Session History Curator};
    \node[module, below=of $(prompt)!0.5!(llm)$] (state) {System State Manager};
    
    \draw[twowayarrow] (attacker) --  (server);
    \draw[arrow] (server) -- (filter);
    \draw[internal] (filter) -- (prompt);
    \draw[internal] (prompt) -- (llm);
    \draw[arrow] (filter) -- (responder);
    \draw[arrow] (responder) to (server);
    \draw[twowayarrow] (history.north) -- (prompt.south);
    \draw[twowayarrow] (state.north) -- (prompt.south);
    \draw[arrow] (llm.south) -- ++(0,-2.5cm) -| (server.south);

    \node[draw, rounded corners, above left=of llm, xshift=1cm] (legend) {
        \begin{tabular}{@{}ll@{}}
        \multicolumn{2}{c}{\textbf{Legend}} \\[.5em] 
        \tikz\node[actor, scale=0.3, text width=3em]{}; & External Actor \\
        \tikz\node[core, scale=0.3, text width=3em]{}; & Core Component \\
        \tikz\node[module, scale=0.3, text width=3em]{}; & Advanced Module \\
        \end{tabular}
    };
    \end{tikzpicture}
\caption{The Canonical Architecture of an LLM-Powered Honeypot. This diagram synthesizes the architectural patterns that have emerged from recent literature. 
Components are color-coded, with \textcolor{blue!80!black}{blue nodes} denoting core, needed components, and \textcolor{violet!80!black}{purple nodes} advanced/optional components introduced in the literature.
An \textcolor{red!80!black}{\textbf{Attacker}} connects to the \textcolor{blue!80!black}{\textbf{Attacker-Facing Server}}, which mimics a network service (e.g., SSH). 
Each command is passed to a \textcolor{violet!80!black}{\textbf{Filter/Router}}, which decides if/when to use the LLM, and then passes the command to a \textcolor{violet!80!black}{\textbf{Deterministic Responder}} for simple or cached responses or, for novel interactions, forward it toward an LLM \cite{ragsdale2023designing, Wang2024}. 
The Filter/Router is a critical component for security as it prevents prompt injection, Denial-of-Service (DoS), and Denial-of-Wallet (DoW) attacks \cite{guan2024honeyllm, Wang2024, johnson2024modular}, and lowers the defender's cost by limiting use of the LLM to only when needed. 
The \textcolor{violet!80!black}{\textbf{Prompt Creator}} component constructs a rich, context-aware prompt for the \textcolor{blue!80!black}{\textbf{LLM}}. This can incorporate data from a \textcolor{violet!80!black}{\textbf{System State Manager}}, which tracks changes to the virtual environment, and a \textcolor{violet!80!black}{\textbf{Session History Curator}}, which intelligently prunes the command history to manage context length \cite{Wang2024}. 
These context-management components seek to increase fidelity of responses and lower both response latency and cost as the LLM processes fewer tokens each call. 
The LLM component itself may represent multiple models or fine-tuned LoRA heads, each specialized for a different protocol. All interactions are logged in the \textbf{Honeypot Data Store} (not depicted) for analysis.}
    \label{fig:architecture}
\end{figure*}
\section{Discussion \& Limitations}\label{sec:discussion}

\subsection{LLM-Honeypot Architecture} \label{sec:architecture}
Our survey indicates a clear convergence toward a multi-component canonical architecture.
Early systems that simply passed commands to a prompted LLM have evolved into more sophisticated designs that add layers for security, efficiency, and fidelity.
This architecture typically includes a pre-LLM filter to block scanners and cache responses, a core LLM engine for generation, and a state manager to track context.
Other components include context pruning modules and prompt generators.
Figure \ref{fig:architecture} provides a general overview of the basic architectural components of an LLM honeypot system that have developed in the literature.

Evidence of the need for these more complex architectures is established, yet no comprehensive, easily configurable architecture that implements all these features exists as open source.
This is a direct engineering gap hindering reproducible research and rapid community progress.

\subsubsection*{LLM Tradeoffs}
A wide variety of LLMs have been evaluated head-to-head by the current literature, and this provides clear guidelines for users.
A fundamental dichotomy exists between using proprietary LLMs, which are generally the largest and most capable models accessible only via API, versus an open-weight model that is usually smaller and weaker but can be run locally.
Research suggests the former requires the user to forfeit inference privacy and entails potentially costly API fees and greater latency, but provides greater deceptiveness and fidelity, focusing development on prompt engineering.
Using the latter---smaller but open-weight models---allows all computation to happen locally with generally faster response times, but entails less accuracy and believability.
When using open-weight models, most research shows worthwhile gains in accuracy and believability after fine-tuning models for specific protocols.
Establishing and managing one's own hardware or cloud infrastructure is the primary cost.

Notably, fine-tuning with Low-Rank Adaptation (LoRA) methods \cite{hu2022lora} promises exciting possibilities for honeypot deployments requiring open-weight LLMs.
As LoRA can be considered a lightweight add-on to the base model, custom LoRA heads can be trained for each protocol to be simulated, essentially providing a fine-tuned model for each protocol while all use the same base model.
This allows a single model to be held in memory and quickly used for different types of inference to enhance the attack surface of the honeypot.

\subsection{LLM-Honeypot Evaluation}
Current research into LLM-based honeypots seeks to resolve what Wang et al. \cite{Wang2024} define as the honeypot ``trilemma'': the simultaneous optimization of flexibility, interaction, and deception. 
To measure progress against these goals, we systematize the literature's disparate metrics into an evaluation tetrad:
(1) \textbf{believability \& deception}, typically measured via human user studies or semantic similarity of simulated versus real outputs;
(2) \textbf{fidelity}, assessed through response timing/latency statistics, Levenshtein distance between generated and ground-truth responses, and command execution success rates;
(3) \textbf{attacker costs \& intelligence}, usually measured  by in-the-wild deployments using metrics such as engagement depth (session length) and the diversity of unique MITRE ATT\&CK patterns elicited; and 
(4) \textbf{defender cost}, comprising token consumption, API latency, and hardware overhead.
We note that very recent work of Aradi et al. \cite{aradi2025metrics} provides a more comprehensive honeypot evaluation framework, including metrics for attack interaction depth (3) and fingerprinting resistance (1), and are recommended to be considered in future evaluations. 

Conceptually, a honeypot succeeds by either wasting adversary resources or generating new threat intelligence, and these gains must be balanced by the defender's cost. 
Consequently, metrics for (3) and (4) are the primary indicators of overall success or ``net security gain.'' 
Believability (1) and fidelity (2) function as metrics that diagnose why a system fails to maintain engagement, as these categories quantify the susceptibility to honeypot fingerprinting vectors. (See Section \ref{sec:honeypot-detection} and Table \ref{tab:honeypot-vectors-llm}.) 

Evaluation and research progress is currently hindered by two fundamental obstacles:

\subsubsection{The Data Desert}\label{sec:data-desert-discussion}
First, real-world deployments are inundated by a ``data desert'' of low-sophistication traffic.
The in-the-wild study by Guan et al. \cite{guan2024honeyllm} demonstrated that even with the most performant LLM tested, average session length (when interaction occurred) only increased from 2.96 to 5.83 commands and 99.2\% of all honeypot activity consisted of simple scanning scripts.
Similarly, Wang et al. \cite{Wang2024} found that in two large datasets, only 0.58\% and 0.048\% of connections resulted in a valid post-login attack session.
This highlights a significant gap: current systems show only marginal improvements in a landscape dominated by low-quality attack noise.
The attack landscape is dominated by activity that is \textit{seemingly} too simplistic to reveal the efficacy of LLM honeypots against sophisticated attackers, and the resulting data is often insufficient for iterative, data-driven development.

These findings raise several critical questions for the development of LLM honeypots: 
``Does low interaction indicate that the attacker has detected the honeypot, or does it merely reflect the simplicity or inflexibility of the attack?'' 
``What level of fidelity is required for different classes of attackers?''
For instance, LLM-driven, context-dependent fidelity may be unnecessary and overly costly for a pre-scripted bot. Conversely, responses that are believable to a human may be identified as artificial by an LLM-powered attacker, and vice versa. (Ayzenshteyn et al. \cite{ayzenshteyn2025cloak}  leverage such discrepancies to expose LLMs). 
Based on the available data, it remains unclear whether the cost of LLM simulation justifies the marginal gain in engagement. 
E.g., a static honeypot populated with detailed, LLM-generated content (honeytokens) \cite{reti2024act} may be a more cost-effective solution for scripted bots, whereas full LLM simulation may only be necessary when attacker actions reach the functional boundaries of static emulators. 
This necessitates the architectural advancements discussed in Section \ref{sec:architecture}, as real-world systems must balance deterministic responses with LLM simulation to optimize the believability, fidelity, and defender/attacker cost-benefit ratio.

\subsubsection{Adversary-Architecture Alignment}
The pre-LLM honeypot era held the belief that low-interaction honeypots will never trick an intelligent human adversary \cite{blackhat_talk_2015}, and, despite the hype, our findings extend this belief to current LLM-driven honeypots.
This stems from fundamental limitations that cannot be solved by realistic simulation; examples persist throughout honeypot history (see Section \ref{sec:honeypot-detection}), such as the inability to convincingly simulate \texttt{vim} or \texttt{htop} outputs without custom architectures, and the even harder challenge of safely downloading and running an adversary's file.
These findings beg the question, ``If humans will not be tricked and bots have limited attack functionality, what is the target adversary of such systems?"
Without a clear attacker model, targeted development of next-generation honeypots is not possible.
Recent work \cite{ilg2025beekeeper} suggests a transition toward an AI-vs-AI paradigm, where the target is not a human, but an autonomous LLM-powered attacker that must be deceived by the honeypot's internal reasoning and output consistency.

To address the lack of a formal attacker model in current literature, we propose a trichotomy of adversaries that maps directly to honeypot architectural requirements and the evaluation tetrad.

\begin{itemize}
    \item \textbf{Tier 1: Prescripted Bots.} These are high-volume, low-sophistication scanners executing fixed, deterministic scripts. For Tier 1, LLM simulation is likely an inefficient use of defender resources (Tetrad 4); they likely are best mitigated by traditional, low-interaction decoys using static, LLM-generated content.
    
    \item \textbf{Tier 2: Autonomous LLM-Agents.} These are sophisticated, AI-driven attackers capable of reasoning, tool-use, and real-time attack strategy adaptation. We argue they are the primary target for LLM-powered honeypots and conjecture deception against Tier 2 requires high  believability and consistency (Tetrads 1 \& 2), but offer the opportunity for wasting ample attacker resources while gaining valuable intelligence (3). 
    
    \item \textbf{Tier 3: Skilled Human Adversaries.} As discussed in Section \ref{sec:honeypot-detection} simulation, LLM or not, is insufficient, and high-interaction environments are likely needed.  
\end{itemize}

\subsection{Automated Threat Intelligence}
The most significant recent breakthrough in honeypot data analysis is the newfound ability to automate the labeling of attacker behavior with MITRE ATT\&CK tactics, largely thanks to fine-tuned language models \cite{boffa2024logprecis}.
This marks the ``Semantic Leap" from processing raw data to generating structured intelligence.
However, this research is still in a nascent stage.
The current state-of-the-art, LogPrécis, is a proof-of-concept, limited by the lack of large-scale, well-labeled datasets.
A clear need for real-world use is a sufficiently capable labeler, which depends on the creation of a larger and more attack-representative dataset.
We also note the inherent difficulty and subjectivity of ATT\&CK mapping~\cite{jiang2025mitre}; perhaps researchers can find a more appropriate taxonomy of honeypot data for aiding cyber defense.
Meanwhile, more advanced concepts, such as using honeypot data to power real-time detection via RAG systems, are emerging but remain exploratory yet promising~\cite{lanka2024}.
\section{Future Research Directions}\label{sec:future-research}
The takeaways from our systematization point to a clear and actionable research roadmap to address the gaps and capitalize on new synergies.

\subsection{Redefining the Target Adversary}
Explicitly, the appropriate target for these systems is not skilled humans, but sophisticated automated attackers (LLM-powered bots)~\cite{fang2024llm, glazunov2024project}.
Historically, automated attacks have been pre-scripted and often repeated.
However, recent research demonstrates that LLMs can be used to create a new class of powerful, autonomous attacker capable of reasoning, using tools, and adapting its strategy mid-attack~\cite{fang2024llm, glazunov2024project, xu2025forewarnedforearmedsurveylarge}.
This marks a fundamental shift in the nature of automated threats, moving from brittle scripts to intelligent agents.
Further, there is now evidence that LLM-powered attack agents are being used in the wild~\cite{reworr2025llmagenthoneypotmonitoring}.
This emerging threat creates an urgent imperative for a new generation of high-fidelity honeypots with increased dynamism and believability.

\subsubsection{The Adversarial Research Ecosystem}
Moreover, the emergence of intelligent and automated attackers presents an unprecedented opportunity for honeypot research.
We therefore propose the development of a dynamic adversarial research ecosystem, a contained platform where LLM-powered attackers and honeypots can be pitted against each other for continuous, co-evolving study.
Such an ecosystem promises to solve the ``data desert" problem by generating high-quality attack data at scale and provide a testbed for developing meaningful metrics for a honeypot's two fundamental goals, occupying attacker resources and gathering novel threat intelligence.
Notably, labeled attacks could be generated by such attackers, which would permit automated attack labeling techniques (e.g., Boffa et al. \cite{boffa2022towards}) to advance from proof-of-concept to viable honeypot components.

\subsection{Architectural Needs}
Future architectural work should focus on building open-source, modular honeynet frameworks to build complex environments.
Significant research is also needed into the use of Small Language Models (SLMs) for privacy- and cost-constrained deployments.
This includes developing efficient fine-tuning and distillation techniques.
The use of LoRA to enable a single base model to flexibly simulate multiple protocols or perform internal tasks is a particularly promising direction for enhancing capability with minimal overhead.

\subsubsection{Unsolved Technical Challenges in Simulation}
While modular architectures and efficient models provide a path forward, significant research is still needed to overcome the core technical hurdles of simulation.
Our surveys highlight several recurring, unsolved problems that future work must address to enhance realism:
\begin{itemize}
    \item Dynamic \& Interactive Commands: Honeypots consistently fail to convincingly simulate commands with continuously updating output (e.g., \texttt{top}) or those that require an interactive, stateful application environment (e.g., \texttt{vim})~\cite{weber2024don}.
    \item {Complex Commands:} LLMs often struggle to correctly parse and execute long, chained, or complex shell commands, leading to unrealistic error messages or incorrect behavior~\cite{weber2024don}.
    \item {Outbound Network Actions:} Simulating actions that require real outbound network connectivity, such as downloading a file with \texttt{wget} or connecting to a C\&C server, remains a fundamental limitation that can quickly expose the honeypot~\cite{ragsdale2023designing}.
\end{itemize}

\subsubsection{Hardening the Honeypot Architecture}
A driver for the more advanced honeypot architectures has been securing the honeypot itself against prompt injections, flooding attacks, or context overflow.
Continued research to identify and mitigate these inherent vulnerabilities will be needed.

\subsubsection{Potential Architectural Directions}
To extend the canonical LLM-honeypot model (Figure \ref{fig:architecture}) and address ongoing challenges in realism and adaptability, we identify avenues for research into alternative architectures, grounded in recent LLM advancements~\mbox{\cite{vaswani2017attention, fan2023bibliometric}}. These directions merit investigation to develop efficient autonomous honeypots that can engage sophisticated automated attackers~\mbox{\cite{heckel2024countering, reworr2025llmagenthoneypotmonitoring}}, subject to empirical validation in cybersecurity contexts.
\begin{itemize}
    \item Domain-specific tokenizers and RAG/tool-use integrations to improve handling of cybersecurity data and enable dynamic, interactive responses~\cite{jaffal2025ai, qiao2022reasoning, boffa2024logprecis}.
    \item {Hybrid selective state space models (e.g., Mamba {\cite{bae2025hybrid}}) for scalable state management of extended sessions, reducing latency in resource-limited deployments.}
    \item {Hierarchical reasoning models like Sapient's HRM~\mbox{\cite{wang2025hrm}} to facilitate abstracted deception strategies with low data requirements, alleviating the ``data desert."}
    \item {Asynchronous multi-agent frameworks, building on honeypot specific systems {\cite{newsham2025sandman, landolt2025marl}}, for parallel execution of duties such as state tracking and adaptive planning.}
\end{itemize}
Pursuing these avenues could enable  autonomous, self-improving deception systems, potentially mitigating the evaluation paradox and data scarcity issues synthesized in Section \ref{sec:data-desert-discussion}.

\subsubsection{REST/JSON API Proposal}
While this work focuses on shell-based honeypots, the extension of LLM simulation to REST/JSON APIs represents a promising avenue for future research.
As the fundamental backend for mobile and web applications, these APIs constitute a critical attack surface exposed to external threats.
The generative capabilities of LLMs are uniquely suited to mimicking complex APIs by producing dynamic, realistic-looking responses that are necessary to sustain attacker engagement.
This high-fidelity simulation creates an opportunity to capture sophisticated probing for business logic flaws and other vulnerabilities.
We anticipate that such honeypots would observe a broad spectrum of attacks, ranging from credential stuffing and data scraping to command injection attempts aimed at the underlying infrastructure fronted by the HTTP server.

\subsection{Toward Real-Time Threat Detection}
Previous work, e.g. LogPrécis of Boffa et al.  \cite{boffa2024logprecis} demonstrate the use of LLMs for (offline) attack pattern detection in logs. 
Future research should expand to real-time attack pattern identification. 
This use case has interesting implications for architectural choices of the underlying models. Boffa et al. \cite{boffa2024logprecis} use a classifier based on an encoder-only model for this purpose. 
In parallel, significant progress has been made on scaling decoder-only (e.g \cite{llama}) and encoder-decoder (\cite{raffel2020exploring}) model architectures. 
We propose a re-evaluation of threat detection tasks in light of these recent advances. 
In particular, causal (decoder-only) architectures more precisely model real-time detection tasks, since no knowledge of future events (honeypot interactions) can be leveraged, is in contrast to threat detection in offline scenarios. 
Encoder-decoder models likewise more naturally map the problem as a translation task from, e.g., raw session interactions to, e.g., MITRE ATT\&CK labels. 
Future work should study what further benefits and drawbacks the inductive biases of these architectures have on tasks such as this one.

\subsection{The Autonomous Feedback Loop}
We now have the potential to create autonomous systems that learn from their interactions.
The first step is to operationalize attack labeling by creating the large-scale, open-source datasets needed to train production-grade classifiers.
With reliable labeling, two powerful feedback loops become possible: a honeypot-to-SOC loop that automatically feeds intelligence to live defensive tools, and a honeypot-to-honeypot loop that enables self-improvement through automated reconfiguration.

With automated TTP labeling, it becomes possible to create novel metrics that move beyond simple session length.
For example, a metric like information gain, based on the novelty of observed attack sequences compared to historical data, could quantify the honeypot's effectiveness at gathering new intelligence.
These new, quantifiable metrics can then serve as a formal objective function for optimization algorithms (e.g., Reinforcement Learning or prompt optimization techniques).
This provides a clear path toward data-driven methods for establishing stopping criteria (i.e., when a honeypot's information gain diminishes) and for guiding the automated reconfiguration and redeployment of the honeypot to maximize its intelligence-gathering goals.
Leveraging the already-discovered metrics for honeypot success can lead to clear stopping criteria, signaling a need for honeypot reconfiguration, and data-driven methods for how to reconfigure and automatically redeploy the honeypot.
Such a self-adaptive process can ideally speed up the acquisition of threat intelligence.

In this vein, we propose two longer-term research directions.
The first is a federated network of autonomous honeypots for shared threat intelligence and dynamic reconfiguration.
Such an endeavor will entail new challenges, such as developing privacy-preserving techniques to gain adoption.
The second is seeking algorithms for real-time reconfiguration mid-attack.
As an example, given the ability to label an attacker's tactics automatically, real-time prediction of the attacker's next steps can be trained and integrated to assist the LLM honeypot in changing its simulation state to ``steer'' the attacker.
\section{Conclusion}
The integration of LLMs into honeypot systems marks a pivotal moment in the field of cyber deception.
While LLMs were initially seen as a silver bullet for the classic fidelity-risk paradox, our systematization has shown that their immediate impact has been modest.
The true potential of this technology, we argue, is not merely to create slightly more believable decoys, but to confront the next generation of autonomous, intelligent threats.
The rise of LLM-powered attackers creates an urgent need for equally sophisticated, AI-driven defenses.

In this paper, we provided a foundational understanding of this new domain.
We began by creating a taxonomy of the fundamental ways honeypots are detected, framing the core challenges that must be addressed.
We then synthesized the initial literature on LLM-powered honeypots into a canonical architecture and systematized the field's approaches to evaluation.
Finally, we charted the evolution of log analysis, showing a clear progression toward the ultimate goal of automated threat intelligence.

Our key insight is that these distinct research threads—architecture, evaluation, and analysis—converge on a powerful new paradigm: the autonomous, self-improving honeypot that operates in a continuous feedback loop of interaction, analysis, and reconfiguration.
By providing these structured frameworks and a clear research roadmap, we hope to guide the community in building the intelligent, adaptive deception systems necessary to secure the networks of tomorrow.
\section*{Acknowledgments}
This research was funded by 
Vinnova, the Swedish Innovation Agency.

The authors utilized Google Gemini 2.5 Pro \cite{google_gemini} to assist in the preparation of this manuscript.
The initial draft of all content was written entirely by the authors.
This author-drafted text was then provided to Gemini for suggestions on improving wording, grammar, and organization.
Gemini was also used to validate \LaTeX \ syntax and to help identify potential errors in the references.
All changes suggested by the model were manually reviewed by the authors.
The authors retained full editorial control.

\bibliographystyle{ieeetr}
\bibliography{refs}

@techreport{Hutchins2011CyberKillChain,
  author    = {Hutchins, Eric M. and Cloppert, Michael J. and Amin, Rohan M.},
  title     = {Intelligence-Driven Computer Network Defense Informed by Analysis of Adversary Campaigns and Intrusion Kill Chains},
  institution = {Lockheed Martin Corporation},
  year      = {2011},
  type      = {White Paper}
}

@incollection{Kemppainen2018,
  author = {Kemppainen, Simo and Kovanen, Tiina}, 
  title = {Honeypot Utilization for Network Intrusion Detection}, 
  editor = {Lehto, Martti and Neittaanmäki, Pekka}, 
  booktitle = {Cyber Security: Power and Technology},
  publisher = {Springer International Publishing AG},
  year = {2018}, 
  pages = {249-270}, 
  series = {Intelligent Systems, Control and Automation: Science and Engineering}, 
  volume = {93}
}

@inproceedings{spyros2022towards,
  title={Towards continuous enrichment of cyber threat intelligence: a study on a honeypot dataset},
  author={Spyros, Arnolnt and Papoutsis, Angelos and Koritsas, Ilias and Mengidis, Notis and Iliou, Christos and Kavallieros, Dimitris and Tsikrika, Theodora and Vrochidis, Stefanos and Kompatsiaris, Ioannis},
  booktitle={International Conference on Cyber Security and Resilience},
  pages={267--272},
  year={2022},
  organization={{IEEE}}
}

@article{Ghourabi2013CharacterizationOA,
        title={Characterization of attacks from the deployment of honeypot},
        author={A. Ghourabi and T. Abbes and A. Bouhoula},
        year={2013},
        journal={Security Communication Networks},
        doi={10.1002/sec.737}
}

@inproceedings{Owezarski2015ANR,
        title={A Near Real-Time Algorithm for Autonomous Identification and Characterization of Honeypot Attacks},
        author={Philippe Owezarski},
        year={2015},
        month={Apr},
        booktitle={ACM Symposium on Information, Computer and Communications Security (ASIACCS)}
        }

@inproceedings{Ryandy2020XTPOTET,
        title={{XT-Pot: eXposing Threat Category of Honeypot-based attacks}},
        author={Ryandy and Charles Lim and Kalpin Erlangga Silaen},
        year={2020},
        month={sep},
        booktitle={The International Conference on Engineering and Information Technology for Sustainable Industry},
        pages={1--6},
        publisher={ACM},
        address={NY, USA},
        doi={10.1145/3429789.3429868},
        url={https://doi.org/10.1145/3429789.3429868}
        }

@INPROCEEDINGS{Aslan2024Unveiling,
    author = {Aslan, \c{C}a\u{g}r\i\ B. and T\"urk\c{s}anl\i, Eftun and Erkan, R. Emre and \"Ozt\"urk, Mustafa and Akdeniz, C\"uneyt},
  booktitle={2024 17th International Conference on Information Security and Cryptology (ISCTürkiye)}, 
  title={Unveiling Hidden Patterns in T-Pot Honeypot Logs: A Latent Topic Analysis}, 
  year={2024},
  volume={},
  number={},
  pages={1-6},
  doi={10.1109/ISCTrkiye64784.2024.10779210}}

@article{lanka2024,
        title={Intelligent Threat Detection—{AI}-Driven Analysis of Honeypot Data to Counter Cyber Threats},
        author={Phani Lanka and Khushi Gupta and Cihan Varol},
        year={2024},
        journal={Electronics},
        volume={13},
        number={13},
        pages={2465},
        doi={10.3390/electronics13132465}
}

@inproceedings{otal2024llm,
  title={{LLM} Honeypot: Leveraging Large Language Models as Advanced Interactive Honeypot Systems},
  author={Otal, Hakan T and Canbaz, M Abdullah},
  booktitle={2024 IEEE Conference on Communications and Network Security (CNS)},
  pages={1--6},
  year={2024},
  organization={IEEE}
}

@inproceedings{reti2024act,
  title={Act as a honeytoken generator! an investigation into honeytoken generation with large language models},
  author={Reti, Daniel and Becker, Norman and Angeli, Tillmann and Chattopadhyay, Anasuya and Schneider, Daniel and Vollmer, Sebastian and Schotten, Hans D},
  booktitle={Proceedings of the 11th ACM Workshop on Adaptive and Autonomous Cyber Defense},
  pages={1--12},
  year={2024}
}

@article{heckel2024countering,
  title={Countering Autonomous Cyber Threats},
  author={Heckel, Kade M and Weller, Adrian},
  journal={arXiv preprint arXiv:2410.18312},
  year={2024}
}

@misc{BeelzebubHoneypot,
  author = {Beelzebub Labs},
  title = {Beelzebub Honeypot},
  note = {Accessed: April 23, 2025},
  howpublished = {\url{https://beelzebub-honeypot.com/}}
}

@misc{TelekomSecurityTPotLLMHoneypot,
  author       = {{Telekom Security}},
  title        = {{T-Pot Community Edition - {LLM-}Based Honeypots Section}},
  howpublished = {GitHub Repository \url{{https://github.com/telekom-security/tpotce?tab=readme-ov-file#llm-based-honeypots}}},
  year         = {2025},
  month        = {May}, 
  note         = {Accessed on May 8, 2025}
}

@misc{AIDaisyGranny,
  author = {Desmarais, Anna},
  title = {A {British} telecommunications company launched an {AI} ``granny" that will waste scammers' time by rambling on the phone for as long as possible.},
  year = {2024},
  month = {11},
  day = {27},
  note = {Accessed: April 23, 2025},
  url = {https://www.euronews.com/next/2024/11/27/meet-daisy-the-ai-granny-that-wastes-the-time-of-phone-scammers}
}

@article{ragsdale2023designing,
  title={On designing low-risk honeypots using generative pre-trained transformer models with curated inputs},
  author={Ragsdale, Jarrod and Boppana, Rajendra V},
  journal={IEEE Access},
  volume={11},
  pages={117528--117545},
  year={2023},
  publisher={IEEE}
}

@article{bridges2023soar,
title = {Testing {SOAR} tools in use},
journal = {Computers \& Security},
volume = {129},
pages = {103201},
year = {2023},
issn = {0167-4048},
url = {https://www.sciencedirect.com/science/article/pii/S0167404823001116},
author = {Robert A Bridges and Ashley E Rice and Sean Oesch and Jeffrey A Nichols and Cory Watson and Kevin Spakes and Savannah Norem and Mike Huettel and Brian Jewell and Brian Weber and Connor Gannon and Olivia Bizovi and Samuel C Hollifield and Samantha Erwin},
}

@inproceedings{botta2007towards,
  title={Towards understanding {IT} security professionals and their tools},
  author={Botta, David and Werlinger, Rodrigo and Gagn{\'e}, Andr{\'e} and Beznosov, Konstantin and Iverson, Lee and Fels, Sidney and Fisher, Brian},
  booktitle={Proceedings of the 3rd symposium on Usable privacy and security},
  pages={100--111},
  year={2007}
}

@article{werlinger2009integrated,
  title={An integrated view of human, organizational, and technological challenges of IT security management},
  author={Werlinger, Rodrigo and Hawkey, Kirstie and Beznosov, Konstantin},
  journal={Information Management \& Computer Security},
  volume={17},
  number={1},
  pages={4--19},
  year={2009},
  publisher={Emerald Group Publishing Limited}
}

@article{werlinger2010preparation,
  title={Preparation, detection, and analysis: the diagnostic work of IT security incident response},
  author={Werlinger, Rodrigo and Muldner, Kasia and Hawkey, Kirstie and Beznosov, Konstantin},
  journal={Information Management \& Computer Security},
  volume={18},
  number={1},
  pages={26--42},
  year={2010},
  publisher={Emerald Group Publishing Limited}
}

@article{goodall2004work,
  title={The work of intrusion detection: rethinking the role of security analysts},
  author={Goodall, John and Lutters, Wayne and Komlodi, Anita},
  year={2004}
}

@inproceedings{de2011information,
  title={Information needs of system administrators in information technology service factories},
  author={De Souza, Cleidson RB and Pinhanez, Claudio S and Cavalcante, Victor F},
  booktitle={Proceedings of the 5th ACM Symposium on Computer Human Interaction for Management of Information Technology},
  pages={1--10},
  year={2011}
}

@article{bridges2018information,
  title={How do information security workers use host data? A summary of interviews with security analysts},
  author={Bridges, Robert A and Iannacone, Michael D and Goodall, John R and Beaver, Justin M},
  journal={arXiv preprint arXiv:1812.02867},
  year={2018}
}

@inproceedings{cheswick1992evening,
  title={An Evening with {Berferd} in which a cracker is Lured, Endured, and Studied},
  author={Cheswick, Bill},
  booktitle={Proc. Winter {USENIX} Conference, {San Francisco}},
  pages={20--24},
  year={1992}
}

@book{stoll2024cuckoo,
  title={The cuckoo's egg: tracking a spy through the maze of computer espionage},
  author={Stoll, Cliff},
  year={2024},
  publisher={Simon and Schuster}
}

@misc{UnaPibaGeek_honeypots_detection,
  author = {@UnaPibaGeek},
  title = {honeypots-detection ({GitHub} repository)},
  url = {https://github.com/UnaPibaGeek/honeypots-detection},
  note = {\url{https://github.com/UnaPibaGeek/honeypots-detection} Accessed: 11 March 2025}
}

@article{reworr2025llmagenthoneypotmonitoring,
    author = {Reworr and Dmitrii Volkov},
    title={{LLM} Agent Honeypot: Monitoring AI Hacking Agents in the Wild}, 
    journal = {arXiv preprint arXiv:2410.13919 } ,
    year = 2025
}

@article{valeros2023attacker,
  title={Attacker profiling through analysis of attack patterns in geographically distributed honeypots},
  author={Valeros, Veronica and Rigaki, Maria and Garcia, Sebastian},
  journal={arXiv preprint arXiv:2305.01346},
  year={2023}
}

@misc{blackhat_talk_2015,
  author = {Dean Sysman  and  Gadi Evron  and  Itamar Sher},
  title = {{YouTube Video from Black Hat Talk}},
  howpublished = {{BlackHat Conference presentation }\url{https://www.youtube.com/watch?v=HiZdkBAFp7Q}},
  year = {2015},
  month = dec,
  day = {30},
  note = {Accessed: 25 March 2025}, 
  url = {https://www.youtube.com/watch?v=HiZdkBAFp7Q}
}

@inproceedings{provos2004virtual,
  author = {Niels Provos},
  title = {A Virtual Honeypot Framework},
  booktitle = {13th {USENIX} Security Symposium},
  year = {2004},
  publisher = {{USENIX} Association}
}

@inproceedings{holz2005detecting,
  title={Detecting honeypots and other suspicious environments},
  author={Holz, Thorsten and Raynal, Frederic},
  booktitle={Proceedings from the sixth annual IEEE SMC information assurance workshop},
  pages={29--36},
  year={2005},
  organization={IEEE}
}

@inproceedings{defibaugh2006network,
  title={Network based detection of virtual environments and low interaction honeypots},
  author={Defibaugh-Chavez, P and Veeraghattam, R and Kannappa, M and Mukkamala, S and Sung, AH},
  booktitle={2006 IEEE Information Assurance Workshop},
  year={2006}
}

@inproceedings{fu2006recognizing,
  title={On recognizing virtual honeypots and countermeasures},
  author={Fu, Xinwen and Yu, Wei and Cheng, Dan and Tan, Xuejun and Streff, Kevin and Graham, Steve},
  booktitle={2006 2nd IEEE International Symposium on Dependable, Autonomic and Secure Computing},
  pages={211--218},
  year={2006},
  organization={IEEE}
}

@inproceedings{mukkamala2007detection,
  title={Detection of virtual environments and low interaction honeypots},
  author={Mukkamala, S and Yendrapalli, K and Basnet, R and Shankarapani, MK and Sung, AH},
  booktitle={2007 IEEE SMC Information Assurance and Security Workshop},
  pages={92--98},
  year={2007},
  organization={IEEE}
}

@article{wang2010honeypot,
  title={Honeypot detection in advanced botnet attacks},
  author={Wang, Ping and Wu, Lei and Cunningham, Ryan and Zou, Cliff C},
  journal={International Journal of Information and Computer Security},
  volume={4},
  number={1},
  pages={30--51},
  year={2010},
  publisher={Inderscience Publishers}
}

@article{lukas2021deep,
  title={Deep generative models to extend active directory graphs with honeypot users},
  author={Lukas, Ondrej and Garcia, Sebastian},
  journal={arXiv preprint arXiv:2109.06180},
  year={2021}
}

@inproceedings{wenda2011honeypot,
  title={A honeypot detection method based on characteristic analysis and environment detection},
  author={Wenda, Deng and Ning, Deng},
  booktitle={2011 International Conference in Electrics, Communication and Automatic Control Proceedings},
  pages={201--206},
  year={2011},
  organization={Springer}
}

@article{hu2022lora,
  title={Lora: Low-rank adaptation of large language models.},
  author={Hu, Edward J and Shen, Yelong and Wallis, Phillip and Allen-Zhu, Zeyuan and Li, Yuanzhi and Wang, Shean and Wang, Lu and Chen, Weizhu and others},
  journal={ICLR},
  volume={1},
  number={2},
  pages={3},
  year={2022}
}

@misc{breaking_honeypots_blackhat_2015,
  author = {Dean Sysman and Gadi Evron and Itamar Sher},
  title = {Breaking Honeypots For Fun And Profit and Itamar Sher},
  howpublished = {\url{https://infocondb.org/con/black-hat/black-hat-usa-2015/breaking-honeypots-for-fun-and-profit}},
  event = {Black Hat USA 2015},
  year = {2015},
  note = {Presentation at Black Hat USA 2015, accessed April 25, 2025}
}

@misc{srinivasa2021gottacatchemall,
      title={Gotta catch 'em all: a Multistage Framework for honeypot fingerprinting}, 
      author={Shreyas Srinivasa and Jens Myrup Pedersen and Emmanouil Vasilomanolakis},
      year={2021},
      eprint={2109.10652},
      archivePrefix={arXiv},
      primaryClass={cs.CR},
      url={https://arxiv.org/abs/2109.10652}, 
}

@misc{censys_honeypots_2023,
  author = {{The Censys Research Team}},
  title = {Unmasking Deception: Navigating Red Herrings and Honeypots},
  url = {https://censys.com/blog/red-herrings-and-honeypots},
  year = {2023},
  note = {Accessed: 2025-04-25}
}

@misc{the_honeynet_project,
  author = {{The Honeypot Project}},
  title = {Projects},
  howpublished = {\url{https://www.honeynet.org/projects/}},
  note = {Accessed: 2025-04-25}
}

@inproceedings{cabral2021advanced,
  title={Advanced {Cowrie} configuration to increase honeypot deceptiveness},
  author={Cabral, Warren Z and Valli, Craig and Sikos, Leslie F and Wakeling, Samuel G},
  booktitle={IFIP International Conference on ICT Systems Security and Privacy Protection},
  pages={317--331},
  year={2021},
  organization={Springer}
}

@inproceedings{cabral2019review,
  title={Review and analysis of {Cowrie} artefacts and their potential to be used deceptively},
  author={Cabral, Warren and Valli, Craig and Sikos, Leslie and Wakeling, Samuel},
  booktitle={2019 International Conference on computational science and computational intelligence },
  pages={166--171},
  year={2019},
  organization={IEEE}
}

@inproceedings{vaswani2017attention,
  author = {Ashish Vaswani and Noam Shazeer and Niki Parmar and Jakob Uszkoreit and Llion Jones and Aidan N Gomez and {\L}ukasz Kaiser and Illia Polosukhin},
  title = {Attention is All you Need},
  booktitle = {Advances in Neural Information Processing Systems},
  editor = {I. Guyon and U. Von Luxburg and S. Bengio and H. Wallach and R. Fergus and S. Vishwanathan and R. Garnett},
  volume = {30},
  year = {2017},
  url = {https://proceedings.neurips.cc/paper_files/paper/2017/file/3f5ee243547dee91fbd053c1c4a845aa-Paper.pdf}
}

@article{fan2023bibliometric,
  author = {Lizhou Fan and Lingyao Li and Zihui Ma and Sanggyu Lee and Huizi Yu and Libby Hemphill},
  title = {A Bibliometric Review of Large Language Models Research from 2017 to 2023},
  journal = {arXiv preprint arXiv:2304.02020},
  year = {2023},
  url = {https://arxiv.org/pdf/2304.02020}
}

@article{qiao2022reasoning,
  author = {Shuofei Qiao and Yixin Ou and Ningyu Zhang and Xiang Chen and Yunzhi Yao and Shumin Deng and Chuanqi Tan and Fei Huang and Huajun Chen},
  title = {Reasoning with Language Model Prompting: A Survey},
  journal = {arXiv preprint arXiv:2212.09597},
  year = {2022},
  url = {https://arxiv.org/pdf/2212.09597}
}

@article{zhao2023survey,
  author = {Wayne Xin Zhao and Kun Zhou and Junyi Li and Tianyi Tang and Xiaolei Wang and Yupeng Hou and Yingqian Min and Beichen Zhang and Junjie Zhang and Zican Dong and Du Yifan and Chen Yang and Yushuo Chen and Zhipeng Chen and Jinhao Jiang and Ruiyang Ren and Yifan Li and Xinyu Tang and Zikang Liu and Peiyu Liu and Jian-Yun Nie and Ji-Rong Wen},
  title = {A Survey of Large Language Models},
  journal = {arXiv preprint arXiv:2303.18223},
  year = {2023},
  url = {https://arxiv.org/pdf/2303.18223}
}

@misc{openai_models,
  author = {OpenAI},
  title = {Models},
  howpublished = {\url{https://platform.openai.com/docs/models}},
  urldate = {2025-04-30}
}

@misc{meta_llama,
  author = {Meta},
  title = {Llama models website},
  howpublished = {\url{https://ai.meta.com/llama/}},
  note = {Accessed 2025-05-15}
}

@misc{google_gemini,
  author = {Google},
  title = {Gemini - meet the everyday {AI} assistant from Google},
  howpublished = {\url{https://gemini.google/}},
  note = {Accessed: 2025-04-30}
}

@misc{whiterabbitneo,
  author = {Kindo},
  title = {{WhiteRabbitNeo}: Offensive Security {Gen-AI} Model},
  howpublished = {\url{https://www.securityweek.com/whiterabbitneo-high-powered-potential-of-uncensored-ai-pentesting-for-attackers-and-defenders/}},
    year = 2024,  
    note = {Accessed: 2025-10-20}
}

@Article{jaffal2025ai,
AUTHOR = {Jaffal, Niveen O. and Alkhanafseh, Mohammed and Mohaisen, David},
TITLE = {Large Language Models in Cybersecurity: A Survey of Applications, Vulnerabilities, and Defense Techniques},
JOURNAL = {AI},
VOLUME = {6},
YEAR = {2025},
NUMBER = {9},
ARTICLE-NUMBER = {216},
URL = {https://www.mdpi.com/2673-2688/6/9/216},
ISSN = {2673-2688},
DOI = {10.3390/ai6090216}
}

@article{gupta2023chatgpt,
  title={From chatgpt to threatgpt: Impact of generative {AI} in cybersecurity and privacy},
  author={Gupta, Maanak and Akiri, CharanKumar and Aryal, Kshitiz and Parker, Eli and Praharaj, Lopamudra},
  journal={{IEEE} Access},
  volume={11},
  pages={80218--80245},
  year={2023},
  publisher={IEEE}
}

@article{weber2024don,
  title={Don’t Stop Believin’: A Unified Evaluation Approach for {LLM} Honeypots},
  author={Weber, Simon B and Feger, Marc and Pilgermann, Michael},
  journal={{IEEE} Access},
  year={2024},
  publisher={{IEEE}}
}

@article{malhotra2025llmhoney,
  title={{LLMHoney}: A Real-Time {SSH} Honeypot with Large Language Model-Driven Dynamic Response Generation},
  author={Malhotra, Pranjay},
  journal={arXiv preprint arXiv:2509.01463},
  year={2025}
}

@article{ilg2025beekeeper,
  title={Beekeeper: Accelerating Honeypot Analysis with {LLM}-driven Feedback},
  author={Ilg, Niclas and Germek, Dominik and Duplys, Paul and Menth, Michael},
  journal={IEEE Access},
  year={2025},
  publisher={IEEE}
}

@inproceedings{safargalieva2025ohra,
  title={{OHRA}: dynamic multi-protocol {LLM}-based cyber deception},
  author={Safargalieva, Anastasia and R{\"u}ffer, Artur and Vasilomanolakis, Emmanouil},
  booktitle={Proceedings of the 30th Nordic Conference on Secure IT Systems (Nordsec 2025)},
  year={2025},
  organization={Springer}
}

@article{jimenez2025design,
  title={Design and Development of an Intelligent {LLM}-based LDAP Honeypot},
  author={Jim{\'e}nez-Rom{\'a}n, Javier and Almenares-Mendoza, Florina and S{\'a}nchez-Maci{\'a}n, Alfonso},
  journal={arXiv preprint arXiv:2509.16682},
  year={2025}
}

@inproceedings{sladic2024llm,
 author = {Muris {Sladić} and Veronica Valeros and Carlos Catania and Sebastian Garcia},
 title = {{LLM} in the shell: Generative honeypots},
 volume = {220},
 pages = {430–435},
 year = {2024},
 publisher = {IEEE},
 booktitle = {2024 IEEE European Symposium on Security and Privacy Workshops {(EuroS\&PW)}}
}

@inproceedings{sladic2025vellmes,
  title={{VelLMes}: A High-Interaction AI-Based Deception Framework},
  author={Sladi{\'c}, Muris and Valeros, Veronica and Catania, Carlos and Garcia, Sebastian},
  booktitle={2025 IEEE European Symposium on Security and Privacy Workshops (EuroS\&PW)},
  pages={671--679},
  year={2025},
  organization={IEEE}
}

@inproceedings{ayzenshteyn2025cloak,
  title={Cloak, Honey, Trap: Proactive Defenses Against $\{$LLM$\}$ Agents},
  author={Ayzenshteyn, Daniel and Weiss, Roy and Mirsky, Yisroel},
  booktitle={34th USENIX Security Symposium (USENIX Security 25)},
  pages={8095--8114},
  year={2025}
}

@INPROCEEDINGS{guan2024honeyllm,
  author={Guan, Chongqi and Cao, Guohong and Zhu, Sencun},
  booktitle={2024 IEEE Conference on Communications and Network Security (CNS)}, 
  title={{HoneyLLM}: Enabling Shell Honeypots with Large Language Models}, 
  year={2024},
  volume={},
  number={},
  pages={1-9},
  doi={10.1109/CNS62487.2024.10735663}, 
  note = {\url{https://www.cse.psu.edu/~sxz16/papers/HoneyGPT.pdf}}
}

@inproceedings{fan2024honeyllm,
  title={{HoneyLLM}: A large language model-powered medium-interaction honeypot},
  author={Fan, Wenjun and Yang, Zichen and Liu, Yuanzhen and Qin, Lang and Liu, Jia},
  booktitle={International Conference on Information and Communications Security},
  pages={253--272},
  year={2024},
  organization={Springer}
}

@article{gobel2009amun,
  title={Amun: A {Python} honeypot},
  author={G{\"o}bel, Jan Gerrit},
  journal={Technical Report, University of Mannheim, Germany \url{https://madoc.bib.uni-mannheim.de/2595/1/amunhoneypot2.pdf}},
  year={2009}, 
note = {Accessed: 2025-10-20}
}

@INPROCEEDINGS{christli2024ai,
  author={Christli, Jason Aljenova and Lim, Charles and Andrew, Yevonnael},
  booktitle={2024 16th International Conference on Information Technology and Electrical Engineering (ICITEE)}, 
  title={AI-Enhanced Honeypots: Leveraging {LLM} for Adaptive Cybersecurity Responses}, 
  year={2024},
  volume={},
  number={},
  pages={451-456},
  doi={10.1109/ICITEE62483.2024.10808265}}

@article{kelly2021comparative,
  title={A comparative analysis of honeypots on different cloud platforms},
  author={Kelly, Christopher and Pitropakis, Nikolaos and Mylonas, Alexios and McKeown, Sean and Buchanan, William J},
  journal={Sensors},
  volume={21},
  number={7},
  pages={2433},
  year={2021},
  publisher={{MDPI}}
}

@misc{karimi2024galah,
 author = {Adel Karimi},
 title = {Galah: An {LLM}-powered web honeypot},
 howpublished = {\url{https://github.com/0x4D31/galah}},
 year = {2024},
 note = {GitHub repository, accessed:2025-10-20}
}

@misc{glazunov2024project,
 author = {Sergei Glazunov and Mark Brand},
 title = {Project naptime: Evaluating offensive security capabilities of large language models},
 howpublished = {\url{https://googleprojectzero.blogspot.com/2024/06/ project-naptime.html}},
 year = {2024},
 note = {Accessed June 2024}
}

@article{ikuomenisan2022systematic,
  title={Systematic review of graphical visual methods in honeypot attack data analysis},
  author={Ikuomenisan, Gbenga and Morgan, Yasser},
  journal={Journal of Information Security},
  volume={13},
  number={4},
  pages={210--243},
  year={2022},
  publisher={Scientific Research Publishing}
}

@INPROCEEDINGS{mehta2021,
  author={Mehta, Sajeel and Pawade, Dipti and Nayyar, Yash and Siddavatam, Irfan and Tiwart, Anoop and Dalvi, Ashwini},
  booktitle={2021 International Conference on Innovative Computing, Intelligent Communication and Smart Electrical Systems (ICSES)}, 
  title={Cowrie Honeypot Data Analysis and Predicting the Directory Traverser Pattern during the Attack}, 
  year={2021},
  volume={},
  number={},
  pages={1-4},
  doi={10.1109/ICSES52305.2021.9633881}}

@article{valli2009visualization,
  title={Visualization of honeypot data using Graphviz and Afterglow},
  author={Valli, Craig},
  year={2009}
}

@inproceedings{fraunholz2017data,
  title={Data mining in long-term honeypot data},
  author={Fraunholz, Daniel and Zimmermann, Marc and Hafner, Alexander and Schotten, Hans D},
  booktitle={International Conference on Data Mining Workshops (ICDMW)},
  pages={649--656},
  year={2017},
  organization={IEEE}
}

@INPROCEEDINGS{Bokkena2024enhancing,
  author={Bokkena, Bhargava},
  booktitle={2024 5th International Conference on Smart Electronics and Communication (ICOSEC)}, 
  title={Enhancing IT Security with {LLM}-Powered Predictive Threat Intelligence}, 
  year={2024},
  volume={},
  number={},
  pages={751-756},
  doi={10.1109/ICOSEC61587.2024.10722712}}

@inproceedings{vasilatos2024llmpot,
  title={{LLMPot}: Dynamically Configured {LLM}-based Honeypot for Industrial Protocol and Physical Process Emulation},
  author={Vasilatos, Christoforos and Mahboobeh, Dunia J and Lamri, Hithem and Alam, Manaar and Maniatakos, Michail},
  booktitle={2025 {IEEE} 10th European Symposium on Security and Privacy (EuroS\&P)},
  pages={963--979},
  year={2025},
  organization={IEEE}
}

@article{boffa2024logprecis,
 author = {Matteo Boffa and Idilio Drago and Marco Mellia and Luca Vassio and Danilo Giordano and Rodolfo Valentim and Zied Ben Houidi},
 title = {LogPrécis: Unleashing language models for automated malicious log analysis},
 journal = {Computers \& Security},
 volume = {141},
 pages = {103805},
 year = {2024},
 publisher = {Elsevier Ltd.}
}

@article{raffel2020exploring,
  title={Exploring the limits of transfer learning with a unified text-to-text transformer},
  author={Raffel, Colin and Shazeer, Noam and Roberts, Adam and Lee, Katherine and Narang, Sharan and Matena, Michael and Zhou, Yanqi and Li, Wei and Liu, Peter J},
  journal={Journal of machine learning research},
  volume={21},
  number={140},
  pages={1--67},
  year={2020}
}

@article{llama,
  author = {Touvron, Hugo and Lavril, Thibaut and Izacard, Gautier and Martinet, Xavier and Lachaux, Marie-Anne and Lacroix, Timothée and Rozière, Baptiste and Goyal, Naman and Hambro, Eric and Azhar, Faisal and Rodriguez, Aurélien and Joulin, Armand and Grave, Edouard and Lample, Guillaume},
  ee = {https://www.wikidata.org/entity/Q117812816},
  journal = {CoRR},
  title = {LLaMA: Open and Efficient Foundation Language Models.},
  url = {http://dblp.uni-trier.de/db/journals/corr/corr2302.html#abs-2302-13971},
  volume = {abs/2302.13971},
  year = 2023
}

@inproceedings{ozkok2024honeypot,
  title={Honeypot's best friend? Investigating ChatGPT's ability to evaluate honeypot logs},
  author={Ozkok, Meryem Berfin and Birinci, Baturay and Cetin, Orcun and Arief, Budi and Hernandez-Castro, Julio},
  booktitle={Proceedings of the 2024 European Interdisciplinary Cybersecurity Conference},
  pages={128--135},
  year={2024}
}

@article{huang2019automatic,
 author = {Cheng Huang and Jiaxuan Han and Xing Zhang and Jiayong Liu},
 title = {Automatic Identification of Honeypot Server Using Machine Learning Techniques},
 journal = {Security and Communication Networks},
 year = {2019}
}

@inproceedings{boffa2022towards,
 author = {M. Boffa and G. Milan and L. Vassio and I. Drago and M. Mellia and Z.B. Houidi},
 title = {Towards {NLP}-based processing of honeypot logs},
 booktitle = {{IEEE} European Symposium on Security and Privacy Workshops ({EuroS\&PW})},
 pages = {314–321},
 year = {2022}
}

@misc{tatoris2025,
  author =       {Reid Tatoris and Harsh Saxena and Luis Miglietti},
  title =        {Trapping misbehaving bots in an {AI} Labyrinth},
  howpublished = {\url{https://blog.cloudflare.com/ai-labyrinth/}},
  year =         {2025},
  month =        {3},
  day =          {19},
  note =         {Accessed: 2025-03-24},
}

@article{daniel2025labeling,
AUTHOR = {Daniel, Nir and Kaiser, Florian Klaus and Giladi, Shay and Sharabi, Sapir and Moyal, Raz and Shpolyansky, Shalev and Murillo, Andres and Elyashar, Aviad and Puzis, Rami},
TITLE = {Labeling Network Intrusion Detection System ({NIDS}) Rules with {MITRE ATT\&CK} Techniques: Machine Learning vs. Large Language Models},
JOURNAL = {Big Data and Cognitive Computing},
VOLUME = {9},
YEAR = {2025},
NUMBER = {2},
ARTICLE-NUMBER = {23},
URL = {https://www.mdpi.com/2504-2289/9/2/23},
ISSN = {2504-2289},
DOI = {10.3390/bdcc9020023}
}

@misc{mitreattack,
 author = {{MITRE ATT\&CK}},
 title = {Enterprise Tactics},
 howpublished = {\url{https://attack.mitre.org/tactics/enterprise/}},
 year = {2024},
 note = {Accessed 2024-09-03}
}

@misc{Oosterhof2024Cowrie,
 author = {Michel Oosterhof},
 title = {Cowrie {SSH/Telnet} Honeypot},
 howpublished = {\url{https://github.com/cowrie/cowrie}},
 year = {2024},
 note = {Accessed: 2024-09-03}
}

@mastersthesis{badran2025towards,
  title = {Towards Adaptive Web Honeypots, An experimental Implementation using {LLMs}},
  author = {Badran, Mohamad and Niazi, Tareq},
  year = {2025},
  school = {Malmö University},
  address = {Malmö, Sweden},
   url = {https://www.diva-portal.org/smash/get/diva2:1981340/FULLTEXT02}
}

@article{Morishita2019DetectMe,
 author={Shun Morishita and Takuya Hoizumi and Wataru Ueno and Rui Tanabe and Carlos Hernandez Ganan and Michel van Eeten and Katsunari Yoshioka and Tsutomu Matsumoto},
 title={Detect Me If You... Oh Wait. An Internet-Wide View of Self-Revealing Honeypots},
 booktitle={2019 {IFIP/IEEE} Symposium on Integrated Network and Service Management},
 pages={134-143},
 year={2019},
 organization={IEEE}
}

@inproceedings{johnson2024modular,
  title={A Modular Generative Honeypot Shell},
  author={Johnson, Saul and Hassing, Remco and Pijpker, Jeroen and Loves, Rob},
  booktitle={2024 {IEEE} International Conference on Cyber Security and Resilience (CSR)},
  pages={387--394},
  year={2024},
  organization={IEEE}
}

@mastersthesis{Gizzarelli2024,
  title={Honeypot and Generative AI},
  author={Gizzarelli, Enea},
  year={2024},
  school={Politecnico di Torino}
}

@book{Spitzner2002,
  author={Lance Spitzner},
  title={Honeypots: Tracking Hackers},
  publisher={Addison-Wesley Longman Publishing Co., Inc.},
  year={2002},
  address={Boston, MA, USA}
}

@article{cambiaso2023scamming,
  title={Scamming the scammers: Using chatgpt to reply mails for wasting time and resources},
  author={Cambiaso, Enrico and Caviglione, Luca},
  journal={arXiv preprint arXiv:2303.13521},
  year={2023}
}

@inproceedings{dagon2004honeystat,
  title={Honeystat: Local worm detection using honeypots},
  author={Dagon, David and Qin, Xinzhou and Gu, Guofei and Lee, Wenke and Grizzard, Julian and Levine, John and Owen, Henry},
  booktitle={Recent Advances in Intrusion Detection: 7th International Symposium, RAID 2004, Sophia Antipolis, France, September 15-17, 2004. Proceedings 7},
  pages={39--58},
  year={2004},
  organization={Springer}
}

@misc{outpost24_threat_landscape_2023,
  author={Outpost24 Research Team},
  title={Cyber Threat Landscape Study 2023: Outpost24's honeypot findings from over 42 million attacks},
  year={2023},
  url={https://outpost24.com/blog/honeypot-findings-from-over-42-million-attacks/},
  urldate={2025-04-30}
}

@article{holbel2024utilizing,
  title={Utilizing virtualized honeypots for threat hunting, malware analysis, and reporting.},
  author={Holbel, Reilly and Yerby, Johnathan and Smith, Warner},
  journal={Issues in Information Systems},
  volume={25},
  number={1},
  year={2024}
}

@inproceedings{Vetterl2018,
  author={Alexander Vetterl and Richard Clayton},
  title={Bitter Harvest: Systematically Fingerprinting Low- and Medium-interaction Honeypots at Internet Scale},
  booktitle={12th USENIX Workshop on Offensive Technologies},
  organization={USENIX Association},
  year={2018}
}

@INPROCEEDINGS{hu2024my,
  author={Hu, Yuqi and Cheng, Siyu and Ma, Yuanyi and Chen, Shuangwu and Xiao, Fengrui and Zheng, Quan},
  booktitle={2024 9th International Conference on Big Data Analytics (ICBDA)}, 
  title={{MySQL-Pot}: A {LLM-}Based Honeypot for {MySQL} Threat Protection}, 
  year={2024},
  volume={},
  number={},
  pages={227-232},
  doi={10.1109/ICBDA61153.2024.10607309}}

@article{sezgig-boyacı,
title = {{DecoyPot}: A large language model-driven web {API} honeypot for realistic attacker engagement},
journal = {Computers \& Security},
volume = {154},
pages = {104458},
year = {2025},
issn = {0167-4048},
doi = {https://doi.org/10.1016/j.cose.2025.104458},
url = {https://www.sciencedirect.com/science/article/pii/S0167404825001476},
author = {Anıl Sezgin and Aytuğ Boyacı}
}

@article{Thonnard2008,
  author={O. Thonnard and M. Dacier},
  title={A framework for attack patterns’ discovery in honeynet data},
  journal={Digital Investigation},
  volume={5},
  pages={S128--S139},
  year={2008}
}

@article{McKee2023,
  author={F. McKee and D. Noever},
  title={Chatbots in a honeypot world},
  year={2023},
    journal = {arXiv preprint arXiv:2301.03771},
}

@misc{Wang2024,
  author={Z. Wang and J. You and H. Wang and T. Yuan and S. Lv and Y. Wang and L. Sun},
  title={HoneyGPT: Breaking the trilemma in terminal honeypots with large language model},
  year={2024},
  eprint={2406.01882},
  archivePrefix={arXiv}
}

@article{jiang2025mitre,
  title={MITRE ATT\&CK Applications in Cybersecurity and The Way Forward},
  author={Jiang, Yuning and Meng, Qiaoran and Shang, Feiyang and Oo, Nay and Minh, Le Thi Hong and Lim, Hoon Wei and Sikdar, Biplab},
  journal={arXiv preprint arXiv:2502.10825},
  year={2025}
}

@article{guo2025frontier,
  title={Frontier AI's Impact on the Cybersecurity Landscape},
  author={Guo, Wenbo and Potter, Yujin and Shi, Tianneng and Wang, Zhun and Zhang, Andy and Song, Dawn},
  journal={arXiv preprint arXiv:2504.05408},
  year={2025}
}

@article{javadpour2024comprehensive,
  title={A comprehensive survey on cyber deception techniques to improve honeypot performance},
  author={Javadpour, Amir and Ja'fari, Forough and Taleb, Tarik and Shojafar, Mohammad and Benza{\"\i}d, Chafika},
  journal={Computers \& Security},
  volume={140},
  pages={103792},
  year={2024},
  publisher={Elsevier}
}

@article{hasanov2024application,
  title={Application of large language models in cybersecurity: A systematic literature review},
  author={Hasanov, Ismayil and Virtanen, Seppo and Hakkala, Antti and Isoaho, Jouni},
  journal={IEEE Access},
  year={2024},
  publisher={IEEE}
}

@article{hassanin2024comprehensive,
  title={A comprehensive overview of large language models ({LLMs}) for cyber defences: Opportunities and directions},
  author={Hassanin, Mohammed and Moustafa, Nour},
  journal={arXiv preprint arXiv:2405.14487},
  year={2024}
}

@article{fang2024llm,
  title={{LLM} agents can autonomously exploit one-day vulnerabilities},
  author={Fang, Richard and Bindu, Rohan and Gupta, Akul and Kang, Daniel},
  journal={arXiv preprint arXiv:2404.08144},
  year={2024}
}

@article{krawetz2004anti,
  title={Anti-honeypot technology},
  author={Krawetz, Neal},
  journal={IEEE Security \& Privacy},
  volume={2},
  number={1},
  pages={76--79},
  year={2004},
  publisher={IEEE}
}

@misc{xu2025forewarnedforearmedsurveylarge,
      title={Forewarned is Forearmed: A Survey on Large Language Model-based Agents in Autonomous Cyberattacks}, 
      author={Minrui Xu and Jiani Fan and Xinyu Huang and Conghao Zhou and Jiawen Kang and Dusit Niyato and Shiwen Mao and Zhu Han and Xuemin and Shen and Kwok-Yan Lam},
      year={2025},
      eprint={2505.12786},
      archivePrefix={arXiv},
      primaryClass={cs.NI},
      url={https://arxiv.org/abs/2505.12786}, 
}

@misc{bae2025hybrid,
  title        = {Hybrid Architectures for Language Models: Systematic Analysis and Design Insights},
  author       = {Bae, Sangmin and Acun, Bilge and Habeeb, Haroun and Kim, Seungyeon and Lin, Chien-Yu and Luo, Liang and Wang, Junjie and Wu, Carole-Jean},
  year         = {2025},
  month        = {October},
  howpublished = {arXiv preprint arXiv:2510.04800},
  archivePrefix= {arXiv},
  eprint       = {2510.04800},
  primaryClass = {cs.CL},
  url          = {https://arxiv.org/abs/2510.04800},
  note         = {FAIR at Meta and KAIST AI collaboration; submitted October 6, 2025}
}

@misc{wang2025hrm,
      title={Hierarchical Reasoning Model}, 
      author={Guan Wang and Jin Li and Yuhao Sun and Xing Chen and Changling Liu and Yue Wu and Meng Lu and Sen Song and Yasin Abbasi Yadkori},
      year={2025},
      eprint={2506.21734},
      archivePrefix={arXiv},
      primaryClass={cs.AI},
      url={https://arxiv.org/abs/2506.21734}, 
}

@misc{newsham2025sandman,
  title={Inducing Personality in {LLM}-Based Honeypot Agents: Measuring the Effect on Human-Like Agenda Generation}, 
      author={Lewis Newsham and Ryan Hyland and Daniel Prince},
      year={2025},
      eprint={2503.19752},
      archivePrefix={arXiv},
      primaryClass={cs.AI},
      url={https://arxiv.org/abs/2503.19752}, 
}

@misc{landolt2025marl,
  title        = {Multi-Agent Reinforcement Learning in Cybersecurity: From Fundamentals to Applications},
  author       = {Landolt, Christoph R. and Würsch, Christoph and Meier, Roland and Mermoud, Alain and Jang-Jaccard, Julian},
  year         = {2025},
  institution  = {Cyber-Defence Campus, armasuisse Science and Technology},
  howpublished = {arXiv preprint arXiv:2505.19837},
  url          = {https://arxiv.org/pdf/2505.19837},
  note         = {Presented at NATO STO ICMCIS Symposium, Oeiras, Portugal, May 13–14, 2025}
}

@article{aradi2025metrics,
  title={Metrics-Driven Evaluation and Optimization of Honeypots: Toward Standardized Measures of Deception Effectiveness},
  author={Aradi, Zolt{\'a}n and Botty{\'a}n, S{\'a}ndor and Kail, Eszter and Rig{\'o}, Ern{\H{o}} and B{\'a}n{\'a}ti, Anna},
  journal={Acta Polytechnica Hungarica},
  volume={22},
  number={12},
  year={2025}
}

\end{document}